\documentclass[twocolumn,secnumarabic,amssymb, nobibnotes,superscriptaddress,showkeys, aps, prd]{revtex4-2}

\usepackage{graphicx}	
\usepackage{amsmath}	
\usepackage{amssymb}	
\usepackage{hyperref}
\usepackage{subfigure}
\usepackage{romannum}

\newcommand{\diff}{{\text{d}}}

\newcommand{\aap}{    {\it Astron. \& Astrophys.}}

\newcommand{\apjl}{   {\it Astrophys. J. Lett.}}

\newcommand{\mnras}{  {\it Mon. Not. Roy. Astron. Soc.}}

\newcommand{\na}{    {\it New Astronomy}} 

\newcommand{\apjs}{ {\it The Astrophysical Journal Supplement Series}}

\usepackage{color}

\begin{document}

\title{Impact of scale-height derivative on general\\ relativistic slim disks in tidal disruption events}

\author{T. Mageshwaran}%
\affiliation{Department of Space Science and Astronomy, Chungbuk National University, Cheongju 361-763, Korea}

\author{Kimitake Hayasaki}%
\affiliation{Department of Space Science and Astronomy, Chungbuk National University, Cheongju 361-763, Korea}
\affiliation{Harvard-Smithsonian Center for Astrophysics, 60 Garden Street, Cambridge, MA02138, USA}

\email{tmageshwaran2013@gmail.com, tmageshwaran@chungbuk.ac.kr, kimi@chungbuk.ac.kr}
\date{\today}%

\begin{abstract}
We construct a numerical model of steady-state, general relativistic (GR) super-Eddington accretion flows in an optically thick, advection-dominated regime, motivated by tidal disruption events wherein super-Eddington accretion assumes a pivotal role. Our model takes into account the loss of angular momentum due to radiation and the scale-height derivative in the basic equations of the GR slim disk. For comparison purposes, we also provide a new analytical solution for a radiation-pressure-dominant GR slim disk, which neglects the angular momentum loss due to radiation and the scale-height derivative. We find that the radiation pressure enhances by incorporating the scale height derivative into the basic equations. As a result, the surface density near the disk's inner edge decreases, whereas the disk temperature and scale height increase, brightening the disk spectrum in the soft X-ray waveband. Notably, an extremely high mass accretion rate significantly enhances the effect of the scale-height derivative, affecting the entire disk. In contrast, the inclusion of the radiation-driven angular momentum loss only slightly influences the disk surface density and temperature compared with the case of the scale-height derivative inclusion. The X-ray luminosity increases significantly due to scale height derivative for $\dot{M}/\dot{M}_{\rm Edd} \gtrsim 2$. In addition, the increment is higher for the non-spinning black hole than the spinning black hole case, resulting in a one-order of magnitude difference for $\dot{M}/\dot{M}_{\rm Edd}\gtrsim100$. We conclude that incorporating the scale-height derivative into a GR slim disk model is crucial as it impacts the disk structure and its resultant spectrum, particularly on a soft-X-ray waveband. 
\end{abstract}

\maketitle

\section{Introduction} 
\label{sec:intro}
%

A tidal disruption event (TDE) occurs when a star is disrupted by a black hole's tidal gravity that exceeds the star's self-gravity \citep{1976MNRAS.176..633F,1988Natur.333..523R}. The radius below which {tidal disruption} happens is the tidal radius given by $r_{\rm t} \simeq (M/M_{\star})^{1/3}R_{\star}$ \citep{1975Natur.254..295H}, where $M$ is the black hole mass, and $M_{\star}$ and $R_{\star}$ are the stellar mass and radius respectively. The disrupted debris returns to the pericenter following a {Keplerian} orbit with a mass fallback rate {$\dot{M}_{\rm fb} \propto t^{-5/3}$, where $t$ is the time. However,} the time variation of {$\dot{M}_{\rm fb}$ at early} times and the magnitude of $\dot{M}_{\rm fb}$ at the peak depends on the stellar density {\citep{2009MNRAS.392..332L}, stellar rotation \citep{2019ApJ...872..163G}, and stellar orbital eccentricity \citep{2020ApJ...900....3P}.} If the infalling debris loses its angular momentum on a timescale {shorter} than the orbital time of the fallback debris, {the accretion rate can be equivalent to the mass fallback rate. However, the fallback debris interacts between head and tailed parts due to the general relativistic (GR) apsidal motion, and the resultant energy dissipation by stream-stream collisions results in the formation of an accretion disk \citep{2013MNRAS.434..909H,2016MNRAS.461.3760H,2016MNRAS.455.2253B}. The resultant disk can be circular or elliptical depending on the energy dissipation efficiency \citep{2020A&A...642A.111C}. The disk viscosity dominates the subsequent evolution of the formed disk. As the viscous timescale is usually much longer than the orbital time, the mass accretion rate is not likely to equal the mass fallback rate \citep{1990ApJ...351...38C,2009ApJ...700.1047C,2019MNRAS.489..132M}.}

As the ratio of the tidal radius to the black hole horizon is inversely proportional to $M^{2/3}$, an increase in the mass of the black hole results in the tidal disruption radius approaching the event horizon. This highlights the importance of considering GR dynamics for TDEs with higher-mass black holes.
The innermost stable circular orbit, $r_{\rm ISCO}$, decreases for a prograde {black hole spin \citep{1972ApJ...178..347B} so that the impact of the black hole spin on the accretion dynamics, disk spectra, and light curves can be more significant than in the non-spinning black hole case. This motivated \citet{1973blho.conf..343N} to construct a model of a GR steady-state geometrically thin disk with alpha-viscosity. Recently, \citet{2018MNRAS.481.3348B} have constructed a GR model of a time-dependent, optically thick, geometrically thin disk in the context of TDEs. \citet{2020MNRAS.496.1784M} have developed a GR model of a time-dependent, optically thick, geometrically thin disk with a mass fallback rate at the constant outer radius in the case of both full and partial stellar disruptions. These models are gas-pressure dominant, so that should be consistent with the standard disk model when the accretion rate is mildly lower than the Eddington rate in the non-relativistic, steady-state limit \citep{2002apa..book.....F,2021NewA...8301491M}.} The radiation pressure becomes significantly remarkable as the accretion rate exceeds the Eddington rate. However, a radiation-pressure dominant, optically thick disk, where the advective energy transport is not properly treated, is known to result in a Lightman-Eardley thermal instability \citep{1974ApJ...187L...1L}. Therefore, \citet{1988ApJ...332..646A} has introduced the advection cooling in the energy conservation equation of the optically thick disk at super-Eddington accretion rate and has constructed the optically thick, advection-dominant disk in the non-relativistic limit, so-called slim disk model.


A GR steady-state solution of optically thick, advection-dominated accretion flows with alpha-viscosity has been formulated in the Kerr spacetime by \citet{1994ASIC..417..341L,1996ApJ...471..762A,1998ApJ...498..313G,1998MNRAS.297..739B}. \citet{2009ApJS..183..171S} and \citet{2011A&A...527A..17S} have revisited them and, especially, \citet{2009ApJS..183..171S} found a way to solve them efficiently and numerically (hereafter SA09). The SA09 model has been used by \citet{2020ApJ...897...80W,2021ApJ...918...46W,2022ApJ...933...31W} to fit the X-ray spectra of TDEs ASASSN-14li and ASASSN-15oi at different epochs to estimate the black hole mass and spin by using the mass accretion rates as a parameter. In these disk models, the angular momentum loss due to the radiation in the angular momentum conservation equation and the scale-height derivative in the advection energy flux has been neglected for their simple treatment in the basic equations. However, these two quantities are likely to significantly impact the disk structure. The radiation-driven angular momentum loss augments the radial inflow velocity, thereby exerting influence on both surface density and temperature. \citet{2021ApJ...918...46W} examined the effect of the angular momentum loss due to the radiation and found that it was somewhat effective for a low mass accretion rate and also decreases slightly the disk luminosity by reducing viscous stress. However, it is scarcely delineated whether this trend persists at sufficiently elevated mass accretion rates. On the other hand, the scale-height derivative is likely to affect the disk temperature because of changing the disk energy allocation among viscous heating, radiative cooling, and advective cooling. However, little has been known about how the scale-height derivative has an influence on the disk structure and emission. In any case, either term can produce a difference in the disk emission. Even if this difference is small, it can produce a significant luminosity difference in X-rays emitted from the inner edge radius due to the exponential decay of the thermal X-ray spectrum, as the Wein law suggests.

In this paper, we develop a new stationary GR slim disk model by including the two terms (i.e., the angular momentum loss due to the radiation and the scale-height derivative) in the basic equations. Our model adopts the alpha viscosity prescription and the opacity comprising both Thomson electron scattering and Kramer absorption opacity. Based on the model, we especially examine the effect of the newly added scale-height derivative on the disk structure and emission. In Section \ref{rmodel}, we describe the detailed formulation of our GR slim disk model. For comparison purposes, we present a new analytical solution for GR slim disk basic equations without the two terms in Section~\ref{sec:diskanal} and provide full numerical solutions for the basic equations in Section~\ref{sec:diskfull}. The impact of the two terms on the disk structure and spectrum is first described in Section \ref{sec:twoextreme} for two significantly different accretion rate cases, and then how the disk structure and emission depend on the mass accretion rate with and without including the term of the scale-height derivative in Section~\ref{sec:mdotdepend}. Section~\ref{sec:advcooling} describes the importance of advective cooling in the entire disk region. The differences in black hole spin are also explained in each subsection of Section~\ref{results}. We discuss our results in Section~\ref{discuss} and summarize our conclusions in Section~\ref{sac}.

\section{General relativistic stationary slim disk model}
\label{rmodel}
%

In this section, we present the relativistic slim disk equations derived in the Kerr space-time metric. We transform the Kerr metric in the Boyer-Lindquist coordinate to cylindrical coordinate $\{t,~r,~\phi,~z\}$ and the metric on the equatorial plane is limited to  first order in $z/r$. We follow the authors \citep{1996ApJ...471..762A,2011ApJS..195....7X} to get the four velocities ($u^{i}$) in the equatorial plane. The Kerr metric in the cylindrical coordinate and the four velocities are given in the appendix \ref{kmfv}. Following \citet{1996ApJ...471..762A}, the stress-energy tensor is given by

\begin{equation}
T^{\alpha \beta} = \rho u^{\alpha} u^{\beta} + p g^{\alpha \beta} + S^{\alpha \beta} + \tau^{\alpha \beta},
\end{equation}

\noindent where $\rho$ is the density, $p$ is the pressure, $S^{\alpha \beta}$ is the viscous stress tensor, $\tau^{\alpha \beta}= q^{\alpha} u^{\beta} + u^{\alpha} q^{\beta}$ with radiative energy flux $q^{\alpha}$. The mass conservation is given by $\displaystyle{(\rho u^i)_{\rm ;i}} = 0$ and the radial, and angular momentum conservations are given by $\displaystyle{T^{i r}_{; r} = 0}$ and $\displaystyle{\left(T^i_k \xi^k \right)_{; i}=0}$ respectively, where $\xi^k \equiv \delta_{\,\phi}^k$ is the azimuthal Killing vector and $\delta_{\,\phi}^k$ is the Kronecker delta.

We solve the conservation equations for a stationary $\partial_t = 0 $ and axisymmetric $\partial_{\phi} = 0$ relativistic accretion disk with angular momentum loss due to the radiation. We take the vertical flow to zero in our model and obtain the vertically integrated conservation equations. The derivation of the conservation equations are given in appendix \ref{kmfv}, and here, we write them in terms of the dimensionless variables.

In the steady state, the mass conservation equation results in 

\begin{equation}
\dot{M} = - 2 \pi r_{\rm g} c \Sigma \Delta_k^{1/2} \frac{V}{\sqrt{1-V^2}},
\label{eq:mdot}
\end{equation}
where $\Sigma$ is the disk mid-plane surface density, {$r_{\rm g} = G M/c^2$ is the black hole radius}, $V$ is the radial velocity, and $\Delta_k = x^2 -2 x +j^2$ with $x = r/r_{\rm g}$ and dimensionless black hole spin $j = a/r_{\rm g}$ (see also Appendix \ref{kmfv}).
The radial momentum conservation equation is given by
\begin{equation}
\frac{V}{(1-V^2)^2} \frac{\diff V}{\diff x} + \frac{c_{\rm s}^2}{c^2} \frac{1}{P} \frac{\diff P}{\diff x} = \mathcal{A}_1,
\label{eq:rmomeq}
\end{equation}
where $c_{\rm s} = \sqrt{p / \rho}$ is the sound speed with the disk pressure $p$, $P = \int p \,\diff z \approx 2 p H$ is the vertically integrated pressure with the disk scale height $H$, $A_k = x^4 + x^2 j^2 +2 x j^2$, and 
\begin{eqnarray}
\mathcal{A}_1 &=& \frac{r_g}{c^2} \frac{\mathcal{A}}{r (1-V^2)} \nonumber \\
&=& \frac{\gamma_{\rm L}^2 A_k}{x^4 \Delta_k} (x^3-j^2) (\omega - \omega_{\rm K}^{+})(\omega - \omega_{\rm K}^{-}), \\
\omega &=&  \frac{2 j x}{A_k} + \frac{x^3 \Delta_k^{1/2}}{A_k^{3/2}}\frac{\ell}{\gamma_{\rm L}}, \\ 
\omega_{\rm K}^{\pm} &=& \pm \frac{1}{x^{3/2} \pm j}, \\
\gamma_{\rm L}^2 &=& \frac{1}{1-V^2} + \frac{x^2 \ell^2}{A_k},
\label{eq:lorentz}
\end{eqnarray}

\noindent 
where $\mathcal{A}$ is given by equation (\ref{mathcalA}), $\omega$ is the angular velocity with respect to the stationary observer, and $\omega_{\rm K}^{\pm}$ is the angular velocity of a circular orbit in the Kerr metric. The positive and negative signs indicate the prograde and retrograde directions, respectively.

Here, we define the non-dimensional parameter $\ell = \mathcal{L}/(r_{\rm g} c)$ as the disk angular momentum per unit mass ($\mathcal{L}$) normalized by $r_{\rm g} c$. The angular momentum conservation equation of the disk is then given by
\begin{equation}
\frac{\diff (x \bar{S}^{r}_{\phi})}{\diff x} = \frac{1}{2\pi}\dot{M} c\frac{\diff l}{\diff x} - \frac{r_{\rm g}^2}{c} x l Q_{\rm rad},
\label{angeqn}
\end{equation}
where $\bar{S}_{\phi}^r =  -c \nu \Sigma (\Delta_k^{1/2} A_k^{3/2} \gamma_{\rm L}^3 / x^5) (\diff \omega/\diff x)$ {with the turbulent viscosity $\nu$}. We assume that radiative energy flow is along the vertical direction and $Q_{\rm rad}$ {represents} the radiative energy flux. The scale height of the disk using the first order approximation of $z/r$  and the vertical hydrostatic equilibrium is given by \citep{1997ApJ...479..179A}

\begin{equation}
H  = \frac{1}{\sqrt{\zeta(x) }}\frac{c_{\rm s}}{c}\,r_{\rm g},
\label{eq:scalh}
\end{equation}
where $\zeta(x) \equiv [\ell^2 - j^2 (\epsilon^2 - 1)]/(2 x^4)$ and $\epsilon = u_{\rm t}$. 

The total pressure in the disk is given by {$p= p_{\rm rad} + p_{\rm gas}$, where $p_{\rm rad} = a_{\rm rad} T^4/3$} is the radiation pressure with $T$ as the mid-plane temperature of the disk and $a_{\rm rad}$ is the radiation constant, and {$p_{\rm gas} =  k_{\rm B} \rho T/(\mu_{\rm m} m_{\rm p})$} is the gas pressure, where {$\Sigma = 2\rho{H}$} is the density, $m_{\rm p}$ is the mass of a proton, $k_{\rm B}$ is the Boltzmann constant, and $\mu_{\rm m}$ is the mean molecular weight taken to be ionized solar mean molecular weight of $0.65$. The viscous stress in the co-moving rotating frame is obtained using the orthonormal tetrad basis and is given by {$t_{r\phi} = -x^2 \bar{S}_{\phi}^r / (\gamma_{\rm L} A_k^{1/2} \Delta_k^{1/2} r_{\rm g})$} \citep{1994ASIC..417..341L,1995ApJ...450..508R}. {Assuming an alpha viscosity, $t_{r\phi} = -\alpha P$} such that the viscosity is given by

\begin{equation}
\nu = - 2\alpha \frac{r_{\rm g}}{c} \frac{H p}{\Sigma}\frac{x^3}{A_k \gamma_{\rm L}^2} \left(\frac{\diff \omega}{\diff x}\right)^{-1}.
\end{equation}

The energy conservation is given by $Q_{\rm vis} = Q_{\rm adv} + Q_{\rm rad}$, where $Q_{\rm vis}$ is the viscous heating flux, $Q_{\rm adv}$ is the advection cooling flux, and $Q_{\rm rad}$ is the radiative cooling flux. They are given by \citep{2011ApJS..195....7X,2020ApJ...897...80W}

\begin{eqnarray}
Q_{\rm vis} &=& \nu \Sigma  \left(\frac{c}{r_{\rm g}}\right)^2 \frac{\gamma_{\rm L}^4 A_k^2 }{x^6} \left(\frac{\diff \omega}{\diff x}\right)^{2}, 
\label{viseqn}\\
Q_{\rm rad} &=& \frac{\theta}{3}\frac{\sigma T^4}{\kappa \Sigma},
\label{eq:qrad}
\\
Q_{\rm adv} &=& \frac{1}{2\pi} \frac{\dot{M} c_{\rm s}^2}{r_{\rm g}^2 x^2}  \xi(r),
\label{adveqn}
\end{eqnarray}

\noindent respectively, where $\theta = 64$ in this paper which is same as in \citet{2020ApJ...897...80W,2021ApJ...918...46W} whereas $\theta = 32$ in SA09, $\sigma$ is the Stefan-Boltzmann constant, $\kappa=\kappa_{\rm es}+\kappa_{\rm a}$ with $\kappa_{\rm es}=0.34{\rm cm^{-2}g^{-1}}$ and $\kappa_{\rm a}=3.2 \times 10^{22} T^{-7/2} \Sigma / H ~ {\rm cm^{-2}~g^{-1}}$, and $\beta_{\rm gas} = p_{\rm gas} / p$ is the ratio of gas to total pressure. In addition, $\xi(r)$ is defined as \citep{1996ApJ...471..762A}
\begin{eqnarray}
\xi(r) &\equiv& - \frac{T}{c_{\rm s}^2}\frac{\partial s }{ \partial \ln(r)} \nonumber\\
&=& -\frac{4 - 3 \beta_{\rm gas}}{\Gamma_3 - 1} \frac{x}{T} \frac{\diff T}{\diff x} + (4 - 3 \beta_{\rm gas}) \left\{\frac{x}{\Sigma} \frac{\diff \Sigma}{\diff x}- \right. \nonumber\\
&& \left. \frac{x}{H} \frac{\diff H}{\diff x} \right\}, \nonumber \\
\label{eq:xiform}
\end{eqnarray}
where $s$ is the specific entropy and $\Gamma_3$ is the third adiabatic exponent:
\begin{equation}
\Gamma_3 = 1+ \frac{(4 -3 \beta_{\rm gas})(\gamma_{\rm gas} -1)}{\beta_{\rm gas} + 12 (1-\beta_{\rm gas})(\gamma_{\rm gas}-1)}
\nonumber
\end{equation}
with the gaseous specific heat ratio, $\gamma_{\rm gas}$ \citep{1939isss.book.....C}.

Note that the second term on the right-hand side of equation~(\ref{angeqn}) and the term including $\diff H/\diff x$ of equation (\ref{eq:xiform}) show the angular momentum loss due to the radiation and the scale-height derivative, respectively. Both terms have been neglected in the past GR slim disk models (e.g., \citealt{2009ApJS..183..171S}). In the next subsection, for comparison purposes, we provide an analytic solution for the GR basic equations~(\ref{eq:mdot}), (\ref{eq:rmomeq}), (\ref{angeqn}), (\ref{eq:scalh}), and (\ref{viseqn})-(\ref{adveqn}), where we adopt radiation pressure only and neglect the angular momentum loss due to radiation and scale height derivative.

%
\subsection{Analytical solution for a radiation pressure dominant disk without the angular 
momentum loss due to radiation and scale-height derivative}
\label{sec:diskanal}
%

Here we derive an analytical solution for the GR disk with radiation pressure only and no angular momentum loss due to the radiation. 
The angular momentum of the disk is taken to be the circular angular momentum, $\mathcal{L}_{\rm K}=\ell_{\rm K}\,r_{\rm g}c$, where $\ell_{\rm K}$ is given by
\begin{equation}
\ell_{\rm K}= \pm \frac{x^2 \mp 2 j \sqrt{x} + j^2}{x^{3/4} (x^{3/2}-3 \sqrt{x} \pm 2 j)^{1/2}}
\end{equation}
and the positive sign is here for the prograde orbit, and the negative sign is for the retrograde orbit \citep{1972ApJ...178..347B}. For a disk with no angular momentum loss due to the radiation, equation (\ref{angeqn}) reduces to be $\displaystyle{ x \bar{S}_{\phi}^r = \frac{\dot{M} c}{2 \pi} \left[\ell_{\rm K} - \ell_{\rm K,in}\right]}$, where $\ell_{\rm K,in}$ is the angular momentum at the inner radius and the viscous stress is assumed to be zero there. Moreover, by setting $\ell=l_{\rm K}$ at equation~(\ref{eq:lorentz}) and $\omega = \omega_{\rm K}^{+}$, equation (\ref{viseqn}) reduces to
\begin{equation}
Q_{\rm vis} = \frac{3}{4\pi} \frac{G M \dot{M}}{r^3} f(x,j),
\label{qnre}
\end{equation}

\noindent where $f(x,j)$ is defined as a boundary correction factor given by

\begin{multline}
f(x,j) = \left[1 - \frac{\ell_{\rm K,in}}{\ell_{\rm K}}\right] \left[1 - \frac{2 j}{x^{3/2}} + \frac{j^2}{x^2}\right] \\ \left[1 -\frac{3}{x} + \frac{2 j}{x^{3/2}}\right]^{-1} \left[1 + \frac{j}{x^{3/2}}\right]^{-1}.
\end{multline}

\noindent Furthermore, this equation reduces to $Q_{\rm vis} = (3/4\pi)(GM\dot{M}/r^3)[1 - \ell_{\rm K,in}/\ell_{\rm K}]$ at the non-relativistic limit, where $\ell_{\rm K,in}/\ell_{\rm K} = \sqrt{x_{\rm in}/x} = \sqrt{r_{\rm in}/r}$. 

We approximate the advective energy in equation~(\ref{adveqn}) as $Q_{\rm adv}=(1/2\pi) (\dot{M} c_{\rm s}^2 / r^2)$, where $\xi(r)\simeq1$ is assumed \citep{2002apa..book.....F}. Because $\xi(r)$ is a kind of logarithmic entropy gradient and thus a slowly varying quantity, the logarithmic derivative terms, and the corresponding coefficients are of the order of unity except for near the horizon. Therefore, $\xi(r)\simeq1$ is a crude but reasonable assumption for deriving the analytical solution. For a radiation-pressure dominant disk with the Thomson opacity ($\kappa = \kappa_{\rm es}$) and {$p = p_{\rm rad} = a_{\rm rad}T^4/3$}, the hydrostatic equilibrium $c_{\rm s}^2 = p_{\rm rad}/\rho = 2p_{\rm rad}H/\Sigma$ results in $c_{\rm s} = (8 / \theta)(\kappa_{\rm es}r_{\rm g}/c^2)(1 /\sqrt{\zeta(x)}) Q_{\rm rad} $. The energy conservation equation results in 
\begin{equation}
Q_{\rm rad} = Q_{\rm vis} \left[\frac{1}{2} + \sqrt{\frac{1}{4}+ \frac{384}{\theta^2} \left(\frac{\dot{M}}{\eta \dot{M}_{\rm Edd}}\right)^2 \frac{f(x,j)}{x^5 \zeta(x)}} \right]^{-1}.
\label{eq:anqrad}
\end{equation}

\noindent where $\dot{M}_{\rm Edd}=(1/\eta)4\pi{GM}/(\kappa_{\rm es}c)\simeq2.2\times10^{-2}~{M_\odot/\rm yr}~ (\eta/0.1)^{-1} (M / 10^6 M_{\odot})$ is the Eddington accretion rate with the radiative efficiency, $\eta$, and its fiducial value is 0.1. The surface density and the mid-plane temperature are then given by

\begingroup
\allowdisplaybreaks
\begin{eqnarray}
\Sigma &=& \frac{1}{128 \pi} \frac{1}{\alpha}\frac{c^5}{r_{\rm g}^3 \kappa_{\rm es}^2} \frac{\dot{M} \theta^2}{Q_{\rm rad}^2}   \frac{x \zeta(x)\left[\ell_{\rm K} - \ell_{\rm K,in}\right] }{\Delta_k^{1/2} A_k^{1/2} \gamma_{\rm L}}, \label{siganal} \\
T &=&
\left(
\frac{3}{\theta} \frac{\kappa_{\rm es}\Sigma}{\sigma}  Q_{\rm rad}
\right)^{1/4},
\end{eqnarray} 
\endgroup

\noindent respectively.

%
\subsection{Full solutions for the GR disk basic equations}
\label{sec:diskfull}
%

In this section, we derive the complete solution for the GR disk basic equations, including angular momentum loss due to the radiation and the scale height derivative. After some manipulations, we can rewrite the momentum and energy conservation equations in terms of three differential equations of $V$, $\ell$, and $T$ as follows:
\begin{eqnarray}
\frac{\diff V}{\diff x} &=& V(1-V^2) \frac{\mathcal{N}}{\mathcal{D}},
\label{eq:vtd}
\\
\frac{1}{T} \frac{\diff T}{\diff x} &=& \frac{1+\beta_{\rm gas}}{4-3 \beta_{\rm gas}} \left[ \frac{a_4 e_3 - a_3 e_4}{a_3 e_2 + a_2 e_3} + \frac{a_3 e_1 - a_1 e_3}{a_3 e_2 + a_2 e_3} 
\frac{\mathcal{N}}{\mathcal{D}} \right],\nonumber
\\
\label{eq:Ttd} 
\\
\frac{\diff \ell}{\diff x} &=& \frac{a_4 e_2 + a_2 e_4}{a_3 e_2 + a_2 e_3} - \frac{a_2 e_1 + a_1 e_2}{a_3 e_2 + a_2 e_3} 
\frac{\mathcal{N}}{\mathcal{D}},
 \label{eq:ltd}
\end{eqnarray}
where 
\begin{eqnarray}
\frac{\mathcal{N}}{\mathcal{D}}
\equiv 
\frac{r_2(a_3 e_4- a_4 e_3)+r_3 (a_4 e_2 + a_2 e_4) + r_4 (a_3 e_2 + a_2 e_3)}{r_1 (a_3 e_2 + a_2 e_3) + r_2 (a_3 e_1 -a_1 e_3) + r_3 (a_2 e_1 + a_1 e_2)} \nonumber \\
\label{eq:noverd}
\end{eqnarray}
and each component of $r_i$, $a_i$ and $e_i$ $(i=1\sim4)$ are a function of $M,~x,~j,~V,~\ell$ and $T$, and their detailed functional forms are described in Appendix~\ref{compeqn}. 

For the GR disk with no angular momentum loss due to the radiation, \citet{1996ApJ...471..762A} have demonstrated that the flow crosses the horizon with the  $V^2 > c_{\rm s}^2$, whereas the flow is subsonic at a point far from the black hole. At some point, therefore, the flow must pass through a sonic point where $V^2 \approx c_{\rm s}^2$, meaning $\mathcal{N}=0$ and $\mathcal{D} = 0$. This is the condition that a solution for the basic equations of the GR slim disk should satisfy. The SA09 model has handled $\ell_{\rm in}$ as a free parameter and found solutions to satisfy the sonic point condition by varying $\ell_{\rm in}$. However, we cannot solve the angular momentum equation analytically because the term for the angular momentum loss due to radiation is included there.

We integrate the equations (\ref{eq:vtd}), (\ref{eq:Ttd}) and (\ref{eq:ltd}) inwardly starting at the radius $r_{\rm out}$ which represents the size of the accretion disk. We estimate the boundary values assuming $\omega = \omega_{\rm K}^{+}$, advection energy flux $Q_{\rm adv} = 0$, and radiation driven angular momentum loss term, $x \ell Q_{\rm rad} = 0$ (SA09, \citealp{2022ApJ...933...31W,2021ApJ...918...46W}) that results in $ x \bar{S}^{r}_{\phi} = (1/2 \pi)\dot{M} c (\ell - \ell_{\rm in})$. We find a numerical solution with a sonic point by handling  $\ell_{\rm in}$ as a free parameter. Specifically, we can have the following relations by using $ x \bar{S}^{r}_{\phi} = (1/2 \pi)\dot{M} c (\ell - \ell_{\rm in})$, $Q_{\rm adv}=0$, $c_{\rm s}=\sqrt{p/\rho}$, and $\omega=\omega_{\rm K}^{+}$ at the disk outer region:

\begin{widetext}
\begin{eqnarray}
\frac{\theta}{3} \frac{\sigma T^4}{\kappa_{\rm es} + \kappa_a}\frac{1}{\Sigma} &=& \frac{1}{2\pi}
\dot{M}
\left(\frac{c}{r_{\rm g}}\right)^2 \frac{\gamma_{\rm L} A_k^{1/2}}{x^2 \Delta_k^{1/2}} \left(- \frac{\partial \omega_{\rm K}}{\partial x}\right) (\ell - \ell_{\rm in})  , \label{xouteqn1}\\
2\frac{j x}{A_k} + \frac{x^3 \Delta_k^{1/2}}{A_k^{3/2}}\frac{\ell}{\gamma_{\rm L}} &=& \omega_{\rm K}^{+}, \\
\frac{1}{2\pi}\frac{1}{\alpha}\frac{\dot{M}c}{r_{\rm g}} \frac{x (\ell - \ell_{\rm in})}{\Delta_k^{1/2} A_k^{1/2} \gamma_{\rm L}}
&=& \frac{k_{\rm B} T}{\mu_{\rm m} m_{\rm p}} \Sigma + 
\frac{2}{3\sqrt{2\pi}}
a_{\rm rad} T^4
\left( \frac{1}{\alpha}\frac{\dot{M}}{\Sigma} \frac{r_{\rm g}}{c} \right)^{1/2}
\left[\frac{x(\ell - \ell_{\rm in})}{\zeta(x)\gamma_{\rm L}
\Delta_k^{1/2} A_k^{1/2} }\right]^{1/2}.
\label{xouteqn2}
\end{eqnarray}
\end{widetext}

We solve these equations for given $M,~j,~\ell_{\rm in}$ and $\dot{M}$ to obtain $V$, $\ell$ and $T$ at the considered outer radius. 
These quantities work as the boundary condition for equations (\ref{eq:vtd})-(\ref{eq:ltd}). By varying $\ell_{\rm in}$, we find such boundary values as the solution satisfies the sonic point condition. With the given boundary values, we solve equations~(\ref{eq:vtd})-(\ref{eq:ltd}), which includes angular momentum loss due to the radiation and the scale height derivative for the disk region smaller than the outer radius, using the Implicit Runge-Kutta method iteratively over $\ell_{\rm in}$ until the sonic point condition is satisfied and we have a true solution. 

%
\subsection{Disk spectrum calculation}
\label{sec:spectrumcalc}
%

Assuming the black body radiation from the entire disk, the effective temperature is given by $T_{\rm eff} = \left(Q_{\rm rad}/ 2 \sigma\right)^{1/4}$ through the Stefan-Bolzmann law.
The observed spectral flux {density} is given by
\begin{equation}
F_{\nu,\rm obs} = \int_{\Omega} I_{\nu}(T_{\rm eff},~\nu_{\rm obs}) \, \diff \Omega,
\label{specflux}
\end{equation}
where $I_\nu$ is the specific intensity, $\diff \Omega$ is the differential solid angle subtended by the emission point to the observer and is given by $\diff \Omega = \diff S \cos\theta_{\rm los}/ D_{\rm L}^2$, where $S$ is the area of emission, $D_{\rm L}$ is the luminosity distance of the source to the observer and $\theta_{\rm los}$ is the angle between observer line of sight and disk normal vector. In a relativistic formulation, area of disk in $\{r,~\phi\}$ plane is given by $\displaystyle{\diff S = \sqrt{g_{\rm r r} g_{\rm \phi\phi}} \diff r \diff \phi}$, and is $\diff S= r_{\rm g}^2 \sqrt{A_k/\Delta_k} \diff x \diff \phi $, where $A_k = x^4 + x^2 j^2 +2 x j^2$ and $\Delta_k= x^2 -2 x +j^2$. The gravitational redshift effect is included by using the Lorentz invariant $I_{\nu}/\nu^3$ \citep{1979rpa..book.....R} in such a way as $I_{\nu}(\nu_{\rm obs}) = g^3 I_{\nu}(\nu_{\rm em})$, where $\nu_{\rm em}$ is the emitted frequency and $g$ is the redshift factor. The observed flux density is then given by \citep{2018MNRAS.481.3348B}
\begin{equation}
F_{\nu,\rm obs} = \frac{\cos\theta_{\rm los}}{D_{\rm L}^2} \int_\Omega g^3 I_{\nu}\left(T_{\rm eff},~\frac{\nu_{\rm obs}}{g}\right) \,\diff S,
\label{flux}
\end{equation}
where the radial integral range is taken to be from the inner disk radius to the circularization radius $r_{\rm c}=2 r_{\rm t}$. The spectral luminosity is then calculated to be 
\begin{eqnarray}
\nu L_\nu &=& 4 \pi D_{\rm L}^2 \nu_{\rm obs}F_{\nu, \rm obs}, \nonumber \\
 &=& 8\pi^2\cos\theta_{\rm los} r_{\rm g}^2\nu_{\rm obs}\int_{x_{\rm in}}^{x_{\rm c}} 
g^3 I_{\nu}\left(T_{\rm eff},~\frac{\nu_{\rm obs}}{g}\right) \sqrt{\frac{\mathcal{A}_k}{\Delta_k}}\,dx, \nonumber \\
\end{eqnarray} 
where $x_{\rm in}\equiv{r}_{\rm in}/r_{\rm g}$ and $x_{\rm c}\equiv{r}_{\rm c}/r_{\rm g}$, and we adopt for the specific evaluation that $\theta_{\rm los} = 0^{\circ}$ and $g = 1/ u^{\rm t}$, where $u^{\rm t}$ is the time component of the four-velocity (see Appendix~\ref{kmfv} for details).

%
\subsection{Model setup and parameters}
\label{sec:modelsetup}
%

\begin{table}
\caption{
The four models to consider for our GR solutions. Model I incorporates both radiation-driven angular momentum loss and the derivative of the scale height. Model II incorporates radiation-driven angular momentum loss but does not consider the derivative of the scale height. Model III incorporates the derivative of the scale height but does not consider radiation-driven angular momentum loss. Model IV does not incorporate either physics. For these models, we take $\theta = 64$ in equation (\ref{eq:qrad}).}
\label{pcases}
\begin{tabular}{|c|c|c|}
\hline
Model  & Angular momentum  & Scale height  \\
& loss due to radiation & derivative \\
&& \\
\hline 
&& \\
I & \checkmark & \checkmark  \\
&& \\
II & \checkmark & $\times$ \\
&& \\
III & $\times$ & \checkmark \\
&& \\
IV & $\times$ & $\times$ \\
&& \\
 \hline
\end{tabular}
\end{table}

Our model has free parameters that are the black hole mass $M$ and spin $j$, the stellar mass $M_{\star}$ and radius $R_{\star}$, the angular momentum at the inner radius $\ell_{\rm in}$, and the mass accretion rate $\dot{M}$. We adopt the black hole mass as $M = 10^7 M_{\odot}$ with assuming the disruption of a solar-type star, targeting a TDE candidate ASAS-SN 14li \citep{2016MNRAS.455.2918H,2018MNRAS.475.4011B}. The inner and outer disk radii that we adopt are given by the ISCO radius (cf. \cite{1972ApJ...178..347B}) and $r_{\rm out} = 2500~r_{\rm g}$, respectively. For the comparison purpose, we introduce black hole spin parameters and adopt $j =0$ and $0.998$ for them. Regarding $\dot{M}$, we describe it in the following sections.

Moreover, the four scenarios are examined in our GR solutions. As depicted in Table~{\ref{pcases}}, a model that incorporates both the effect of radiation-induced angular momentum loss and the derivative of the scale height (Model I), a model that accounts for radiation-induced angular momentum loss without incorporating the derivative of the scale height (Model II), a model that accounts for the derivative of the scale height without considering radiation-induced angular momentum loss (Model III), and a model that neglects both of these physical effects (Model IV). In the next section, we provide the solutions with the above parameters for the four models represented in Table~\ref{pcases}.

%
\section{Results}
\label{results}
%

In this section, we first delineate the distinctions between Models I through VI for two extreme $\dot{M}$ scenarios. Subsequently, we elucidate the dependence of the mass accretion rate on the disk structure and emission to scrutinize the influence of the scale height derivative more comprehensively. Lastly, we expound upon the significance of advective cooling throughout the entire disk region. We also compare our Model II and Model IV with the earlier works, i.e., SA09 model and models of \citet{2020ApJ...897...80W,2021ApJ...918...46W} for the confirmation purpose of our model validity (see Appendix \ref{Work_comp} for the details).

%
%
\begin{figure*}
\centering
\subfigure[]{\includegraphics[scale = 0.68]{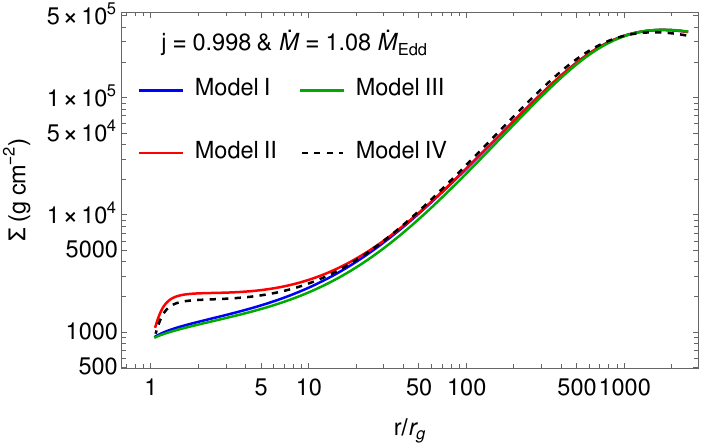}}
\subfigure[]{\includegraphics[scale = 0.66]{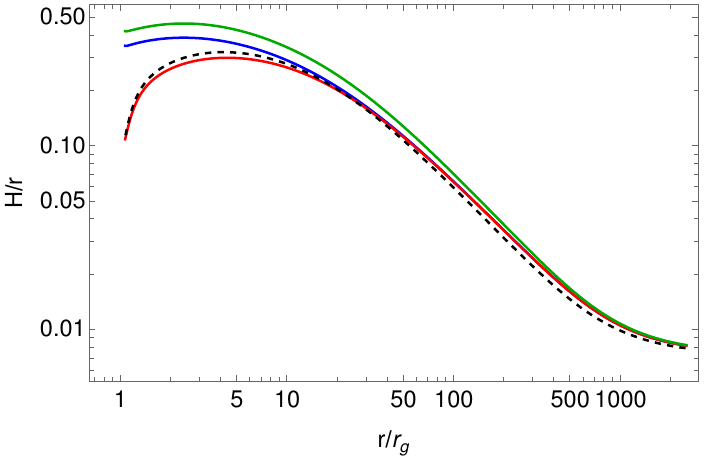}}
\subfigure[]{\includegraphics[scale = 0.66]{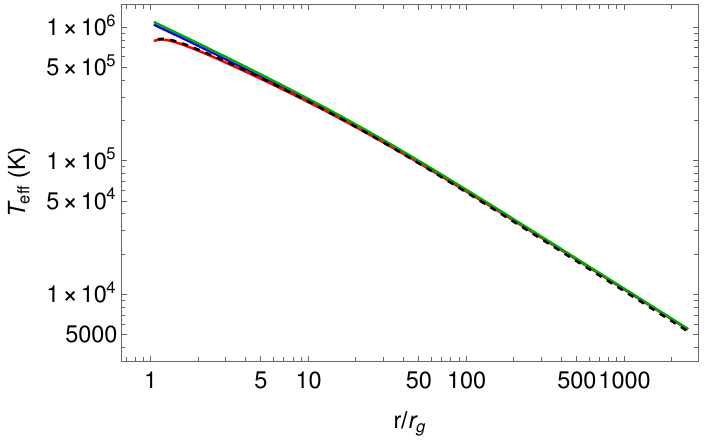}}
\subfigure[]{\includegraphics[scale = 0.6]{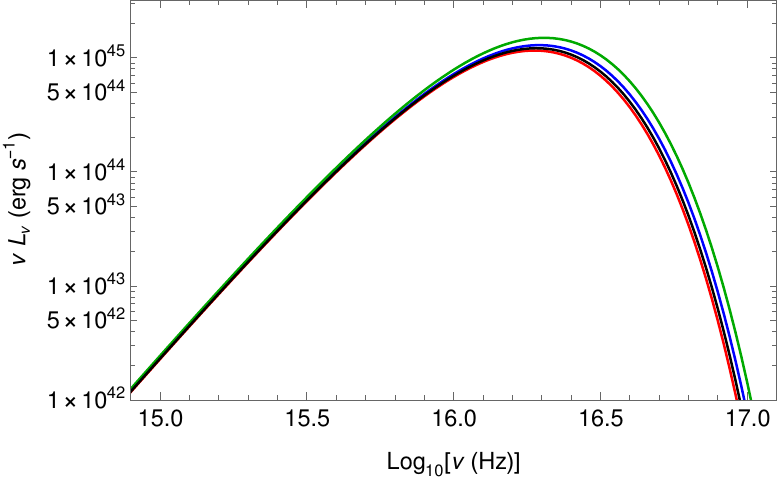}}
\caption{\label{M108j0998} 
The radial dependence of the surface density, scale-height-to-radius ratio, effective temperature, and corresponding disk spectrum with $j = 0.998$ and $\dot{M} = 1.08 \dot{M}_{\rm Edd}$. The blue, red, green and black lines correspond to Models I, II, III, and IV respectively.
}
\end{figure*}
%
%
\begin{figure*}
\centering
\subfigure[]{\includegraphics[scale = 0.6]{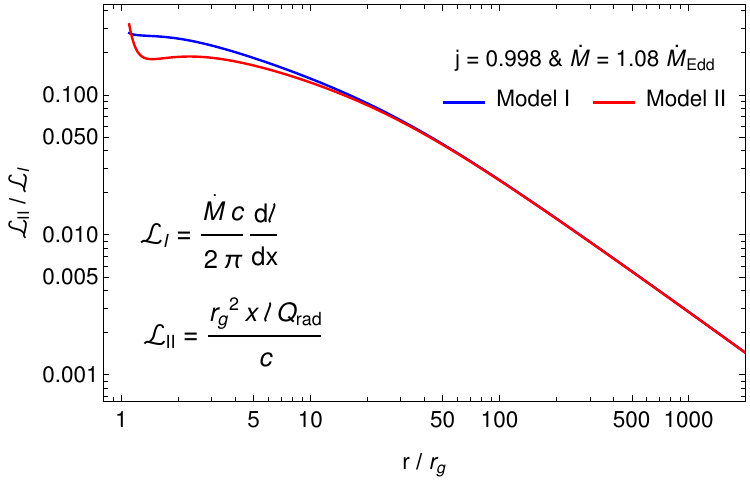}}
\subfigure[]{\includegraphics[scale = 0.6]{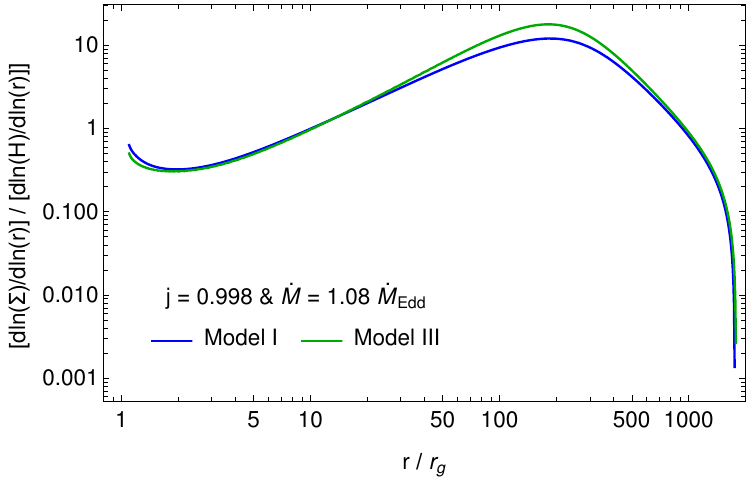}}
\caption{\label{M108j0998_1} (a) Radial dependence of the ratio of $\mathcal{L}_{\rm II} = (r_{\rm g}^2/ c)x \ell Q_{\rm rad} $ and $\mathcal{L}_{\rm I} = (1/2 \pi)\dot{M} c~ \diff \ell / \diff x$ with $j = 0.998$ and $\dot{M} = 1.08\,\dot{M}_{\rm Edd}$. Here $\mathcal{L}_{\rm I}$ and $\mathcal{L}_{\rm II}$ are the first and second terms of the right-hand side of equation~(\ref{angeqn}), respectively. The disk properties are depicted in Figure~\ref{M108j0998}. The blue and red lines correspond to Models I and II, where angular momentum loss due to radiation is included in the model. (b) The radial dependence of the ratio of $\diff \ln(\Sigma)/ \diff \ln(r)$ to $\diff \ln(H)/ \diff \ln(r)$, which is seen in the right-hand side of equation~(\ref{adveqn}). The blue and green lines correspond to Models I and III, which include the scale-height derivative term in solved basic equations. }
\end{figure*}
%
%

%
\subsection{Two distinct mass accretion rate cases}
\label{sec:twoextreme}
%

To clarify the differences between the four models, we choose the following two cases: $\dot{M} = 1.08 \dot{M}_{\rm Edd}$ and $600 \dot{M}_{\rm Edd}$. Note that $\dot{M} = 1.08 \dot{M}_{\rm Edd}$ is a best-fit solution at one epoch for TDE ASAS-SN 14li X-ray spectrum by \citet{2020ApJ...897...80W} and that $\dot{M} = 600 \dot{M}_{\rm Edd}$ is adopted as an extreme end of the mass accretion rate \citep{2020ApJ...897...80W}. Such a high mass accretion rate is considered from the theoretical viewpoint to know how the extreme mass accretion rate impacts the radiation-driven angular momentum loss and scale-height derivative terms on the disk accretion flow.

Figure~\ref{M108j0998} depicts the radial dependence of the surface density, scale-height-to-radius ratio, effective temperature, and disk spectrum with $j = 0.998$ and $\dot{M} = 1.08\,\dot{M}_{\rm Edd}$ for the four models represented in Table~\ref{pcases}. From the figure, we find that the effective temperature and $H/r$ increase, whereas the surface density decreases with the inclusion of the scale height derivative. The scale height derivative influence strongly on the higher energy side of the resultant disk spectrum. On the other hand, the radiation-driven angular momentum loss plays the opposite role of the scale height derivative. In other words, it tends to decrease the effective temperature and $H/r$ and increase the surface density. Overall, both physical terms only affect the inner region of the disk, and the scale-height derivative impacts these physical quantities more significantly than the angular momentum loss due to the radiation.

Panel (a) of Figure~\ref{M108j0998_1} shows the comparison between the angular momentum transfer with and without the radiation-driven angular momentum loss. 
Here we define $\mathcal{L}_{\rm II}\equiv(r_{\rm g}^2/ c)x \ell Q_{\rm rad}$ and $\mathcal{L}_{\rm I}\equiv(1/2 \pi) \dot{M} c~ \diff \ell / \diff x$. These two terms are seen in equation~(\ref{angeqn}), which represents the angular momentum conservation. Panel (b) of Figure~\ref{M108j0998_1} depicts the comparison of the radial derivative between the surface density and the scale height, which are appeared on the right-hand side of equation~(\ref{adveqn}). Note that we confirm that the radial derivative of the disk temperature is comparable with the other two radial derivative terms for all the models. From the two panels, we note that both terms are important in the inner part of the disk but insignificant in the outer region of the disk. Specifically, it is clear from panel (a) that the radiation-driven angular momentum loss is less important and the scale-height derivative is more important in the inner region of the disk.

%
%
\begin{figure*}
\centering
\subfigure[]{\includegraphics[scale = 0.7]{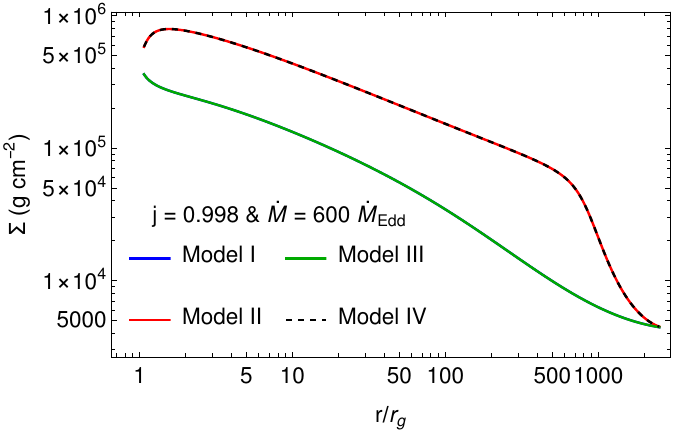}}
\subfigure[]{\includegraphics[scale = 0.64]{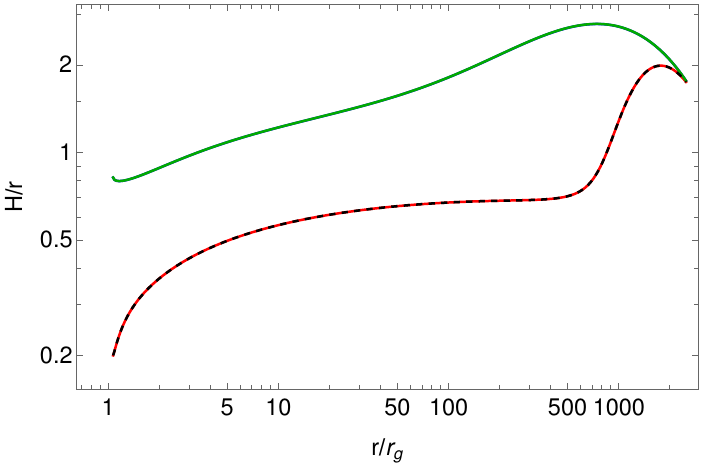}}
\subfigure[]{\includegraphics[scale = 0.7]{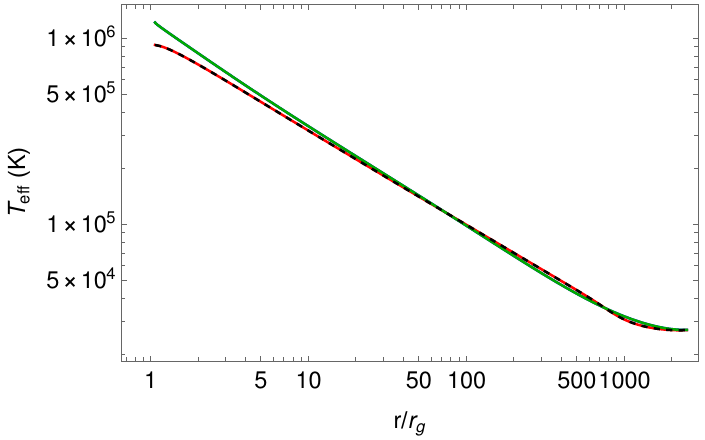}}
\subfigure[]{\includegraphics[scale = 0.6]{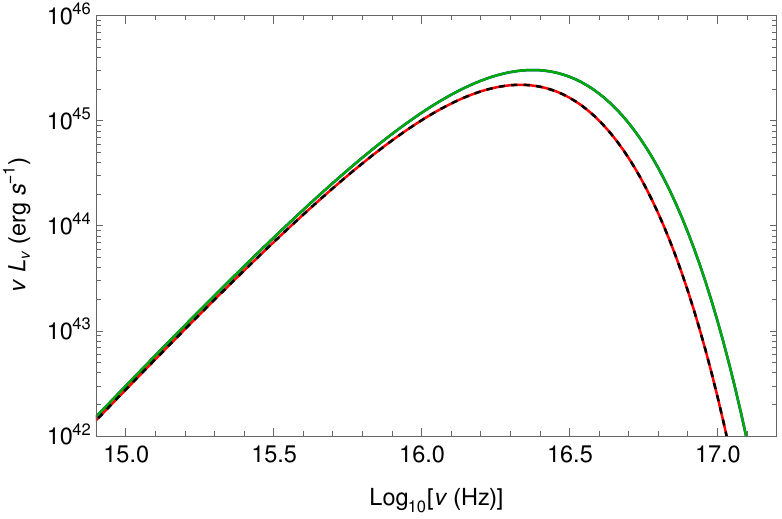}}
\caption{\label{M600j0998} The same format as Figure~\ref{M108j0998} but for $\dot{M} = 600\,\dot{M}_{\rm Edd}$.
}
\end{figure*}
%
%
\begin{figure*}
\centering
\subfigure[]{\includegraphics[scale = 0.7]{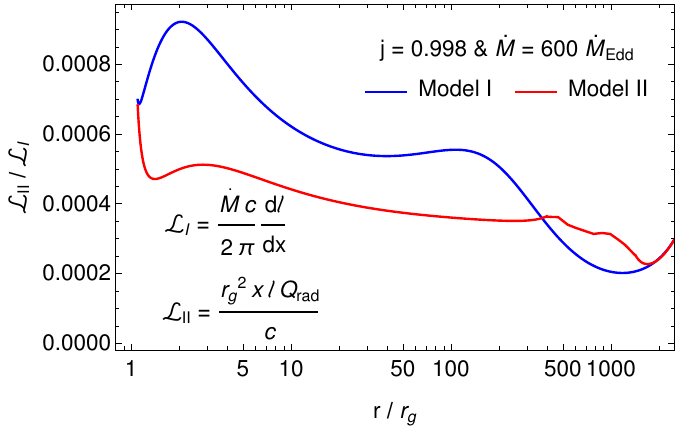}}
\subfigure[]{\includegraphics[scale = 0.67]{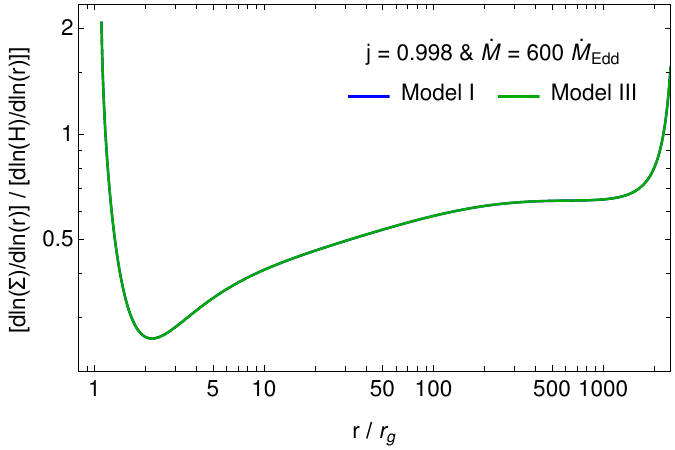}}
\caption{\label{M600j0998_1} The same format as Figure~\ref{M108j0998_1} but for $\dot{M} = 600\,\dot{M}_{\rm Edd}$.}
\end{figure*}

Figure~\ref{M600j0998} and \ref{M600j0998_1} represent the radial dependence of the same physical quantities as in Figure~\ref{M108j0998} and \ref{M108j0998_1}, respectively, but for $\dot{M}=600\,\dot{M}_{\rm Edd}$. From the figure, we find that the effect of the scale-height derivative on these quantities is much more remarkable in the entire region of the disk than the $\dot{M} = 1.08 \dot{M}_{\rm Edd}$ case. This suggests the effect of the scale height derivative is more effective as the mass accretion rate is higher. Panel (a) of Figure~\ref{M600j0998_1} gives a consistent result with this trend. In addition, note that the radiation-driven angular momentum loss more inefficiently works on the disk structure even if the mass accretion rate significantly increases. This property is seen in panel (b) of Figure~\ref{M600j0998_1} because both lines are completely overlapped there.

%
%
\begin{figure*}
\centering
\subfigure[]{\includegraphics[scale = 0.73]{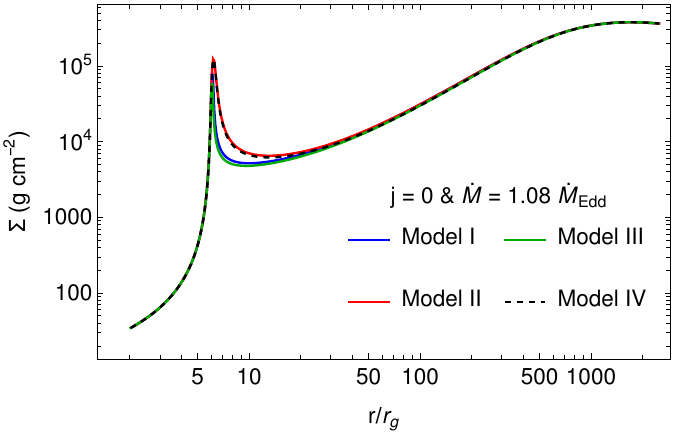}}
\subfigure[]{\includegraphics[scale = 0.72]{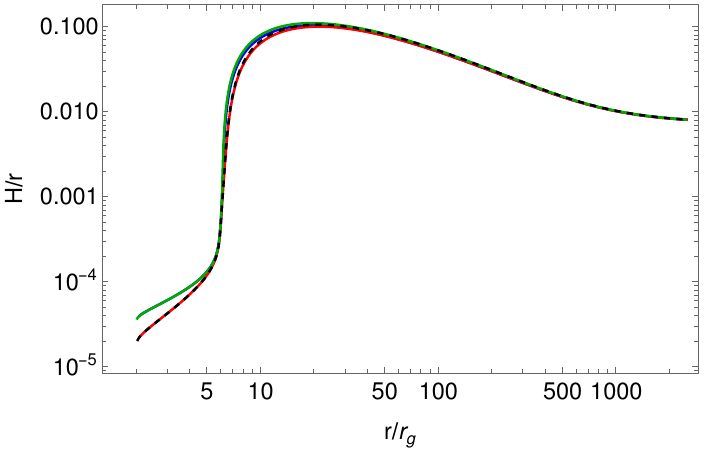}}
\subfigure[]{\includegraphics[scale = 0.73]{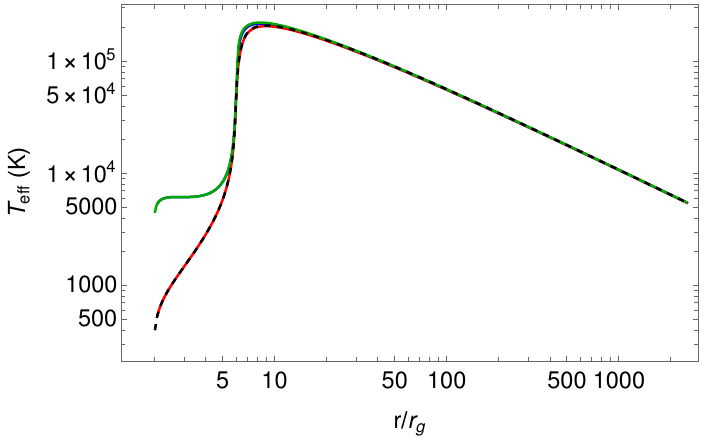}}
\subfigure[]{\includegraphics[scale = 0.65]{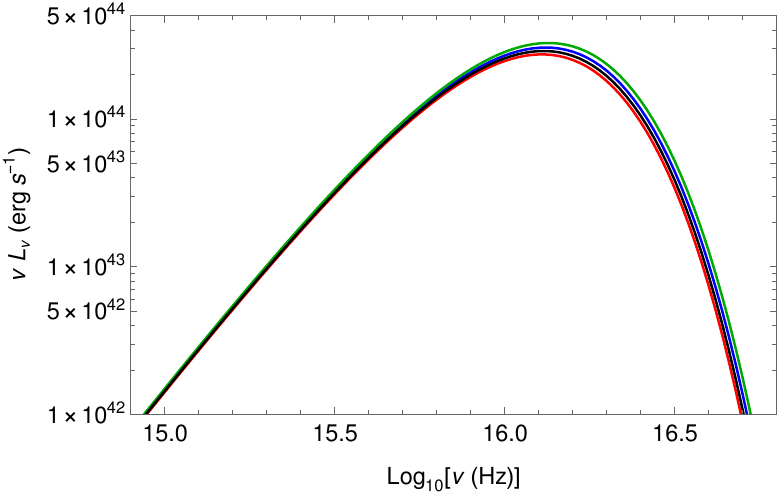}}
\caption{\label{M108j0} The same format as Figure~\ref{M108j0998} but for $j=0$.}
\end{figure*}
%
%
\begin{figure*}
\centering
\subfigure[]{\includegraphics[scale = 0.7]{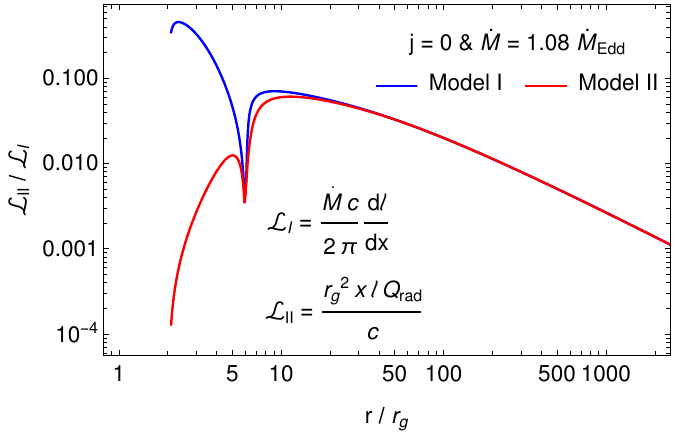}}
\subfigure[]{\includegraphics[scale = 0.7]{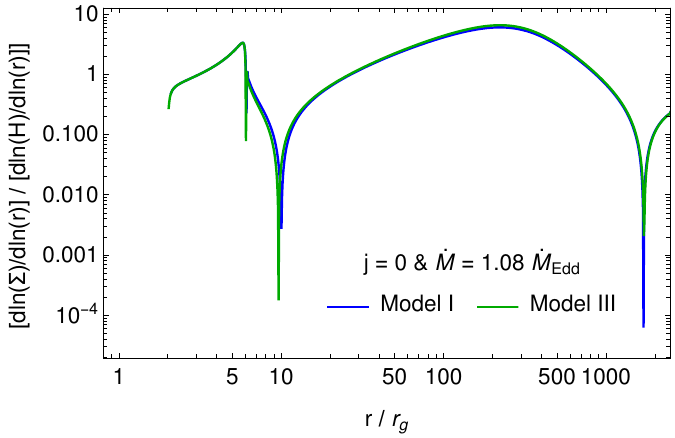}}
\caption{The same format as Figure~\ref{M108j0998_1} but for $j=0$. }
\label{M108j0_1}
\end{figure*}

Figure~\ref{M108j0} displays the radial dependence of the same physical quantities as in Figure~\ref{M108j0998}, except that $j=0$. It is noted from the figure that the surface density profile differs remarkably from the $j=0.998$ case. Moreover, only this model represents the sharp change near the inner radius in the radial surface density profile. Let us explain the cause in more detail below. We confirm that the radiation pressure dominates the gas pressure near inner radii for all the models, except that $j = 0$ and $\dot{M}=1.08\,\dot{M}_{\rm Edd}$, where the dominant pressure transits from radiation to gas pressure at the inner disk region. Since the analytical solution of the surface density, as seen in equation (17), indicates that $\Sigma \propto (\ell - \ell_{\rm in})/Q_{\rm rad}^2 \propto  1 / (\ell - \ell_{\rm in})$, the surface density increases rapidly near the inner radius. On the other hand, the Novikov-Thorne solution has shown that the surface density goes to zero at the inner radius for a gas pressure-dominant case \citep{1973blho.conf..343N}. Combining the two solutions suggests that when the disk transit from radiation to gas pressure near the inner disk radius, the surface density changes from the case where it increases to the case where it decreases sharply. This rapid change creates a hump near the inner disk radius. If the disk does not make such a transition, the surface density changes monotonically, as seen in all other models. In addition, the analytical solution of the radiation pressure-dominated disk assumes a zero viscous stress at the inner radius, whereas the full solution has no such assumption, and the boundary condition is set only at the outer radius. Moreover, the analytical solution has no angular momentum loss due to radiation, but our full solution includes it. These two factors can also produce unusual behavior near the inner radius. For example, the $\diff\Sigma/\diff r$ is zero at certain points where surface density is locally flat, causing the ratio of surface density derivative to scale height derivative near zero at those points. This produces the two sharp droppings as seen in panel (b) of Figure~\ref{M108j0_1}.

\begin{figure*}
\centering
\subfigure[]{\includegraphics[scale = 0.75]{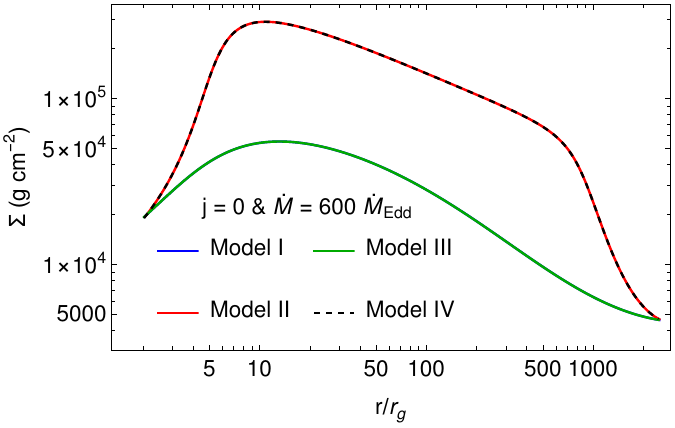}}
\subfigure[]{\includegraphics[scale = 0.71]{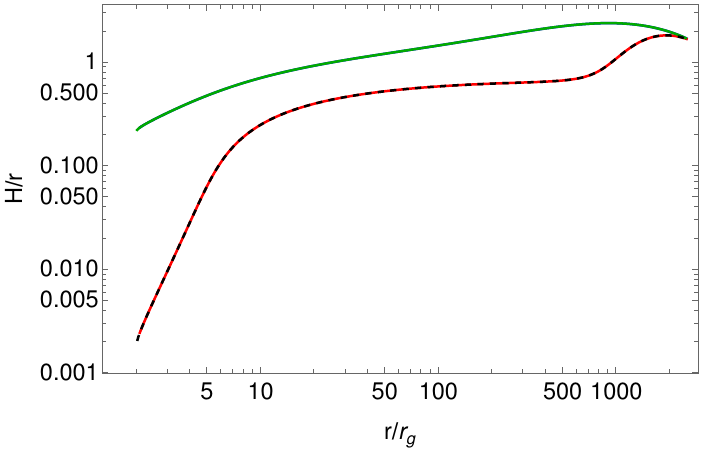}}
\subfigure[]{\includegraphics[scale = 0.74]{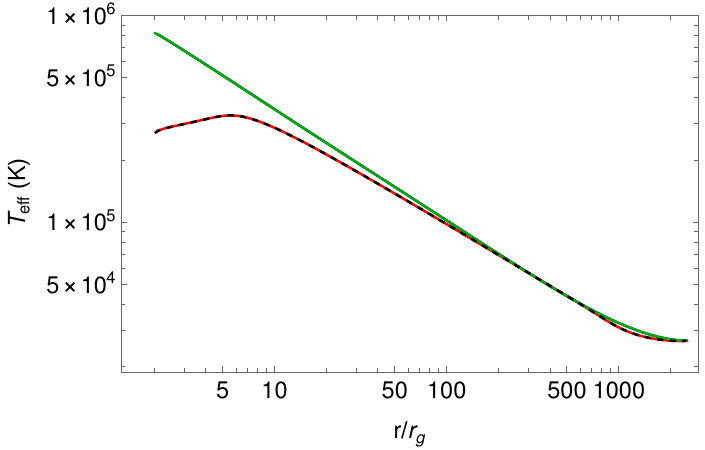}}
\subfigure[]{\includegraphics[scale = 0.64]{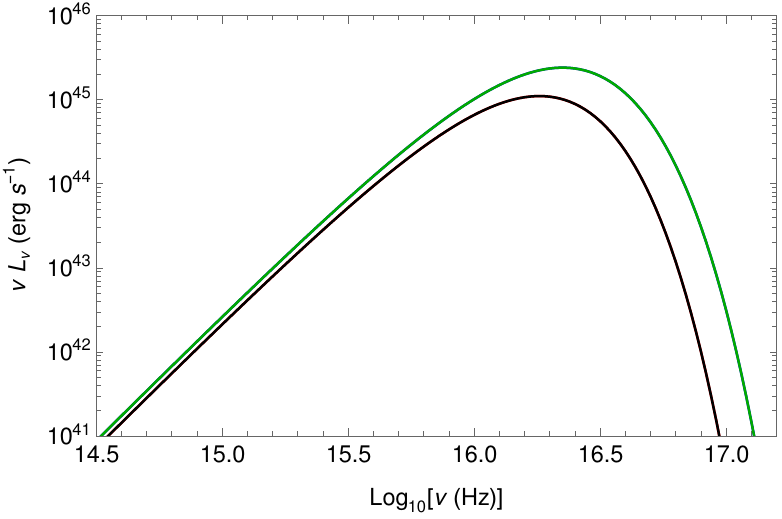}}
\caption{\label{M600j0} The same format as Figure~\ref{M600j0998} but for $j=0$.}
\end{figure*}

\begin{figure*}
\centering
\subfigure[]{\includegraphics[scale = 0.7]{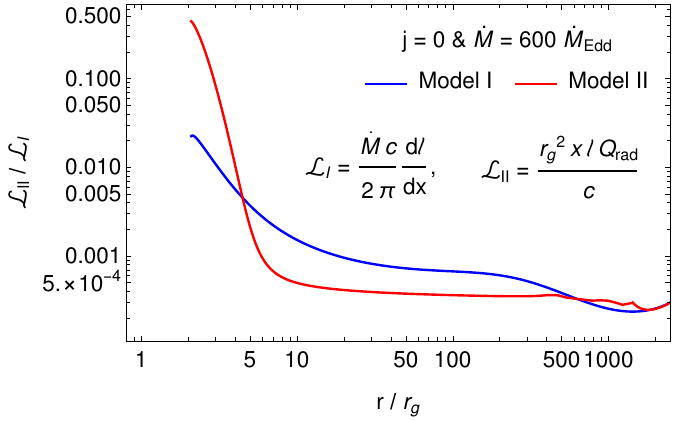}}
\subfigure[]{\includegraphics[scale = 0.7]{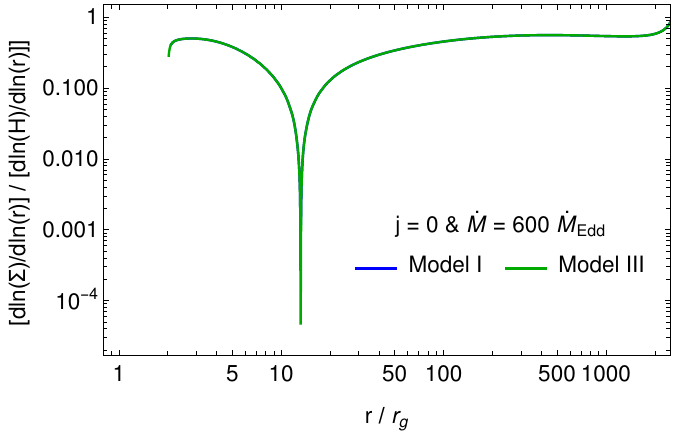}}
\caption{\label{M600j0_1} The same format as Figure~\ref{M600j0998_1} but for $j=0$.}
\end{figure*}

Figure~\ref{M600j0} depicts the radial dependence of the same physical quantities as in Figure~\ref{M600j0998}, except that $j=0$. 
It is noted from the figure that in comparison with $j=0$ and $1.08\,\dot{M}_{\rm Edd}$ case, the effective temperature is much higher, and the resultant peak of the disk spectrum is around $10^{16}\,{\rm Hz}$. While the surface density decreases due to the scale-height derivative, the $H/r$ and effective temperature is higher. 
Panel (a) and (b) of Figure~\ref{M600j0_1} represent quantitatively how important the radiation-driven angular momentum loss and the scale-height derivative terms are the basic equations. We find that the effect of the angular momentum loss due to radiation is negligibly small for both with and without scale-height derivative models as the mass accretion rate increases. Also, the ratio of surface density derivative $\diff\ln\Sigma/\diff\ln{r}$ and scale-height derivative $\diff\ln{H}/\diff\ln{r}$ indicates that the scale-height derivative is important over the entire disk. These tendencies are similar to the spinning black hole case that is depicted in Figure~\ref{M600j0998}. 

Overall, the angular momentum loss due to radiation hardly has an influence on the disk structure and tends to be more insignificant with an increase in the mass accretion rate. The disk luminosity decreases by the inclusion of radiation-driven angular momentum loss, and the decrease in luminosity is negligible for a high mass accretion rate. This result is consistent with \citet{2021ApJ...918...46W} in that they demonstrated that the effect of the radiation-driven angular momentum loss is significant for a lower mass accretion rate. In contrast, the scale height derivative included in the advection energy flux term significantly impacts the disk structure. The effect spreads near the inner edge of the disk to the entire disk as the mass accretion rate increases. The disk spectrum is more luminous due to the inclusion of the scale height derivative, and its increment is higher as the mass accretion rate increases. The luminosity difference is more remarkable due to Wein's law exponential decay of the spectrum around the soft-X-ray waveband of $10^{16-17}$ Hz.

%
\subsection{$\dot{M}$-dependence on the disk structure and emission}
\label{sec:mdotdepend}
%

In Section \ref{sec:twoextreme}, we investigate the effect of the two terms on the disk structure for two extreme cases of mass accretion rate. We find there that the radiation-driven angular momentum loss has a weak impact on the disk structure for both low and high-mass accretion rate cases. In contrast, the impact on the accretion flow due to scale height derivative significantly increases with the mass accretion rate. Here, therefore, we examine in more detail how the disk structure and spectral luminosity evolve with mass accretion rate under the effect of scale height derivative (Model III) and no scale height derivative (Model IV).

\begin{figure}
\centering
\subfigure[]{\includegraphics[scale = 0.67]{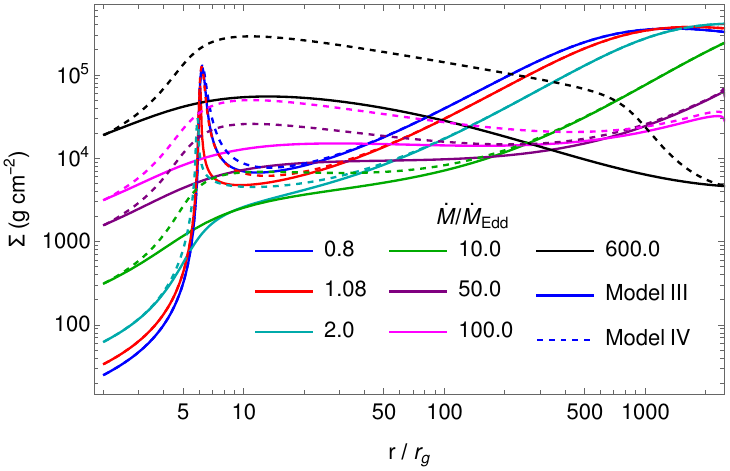}}
\subfigure[]{\includegraphics[scale = 0.67]{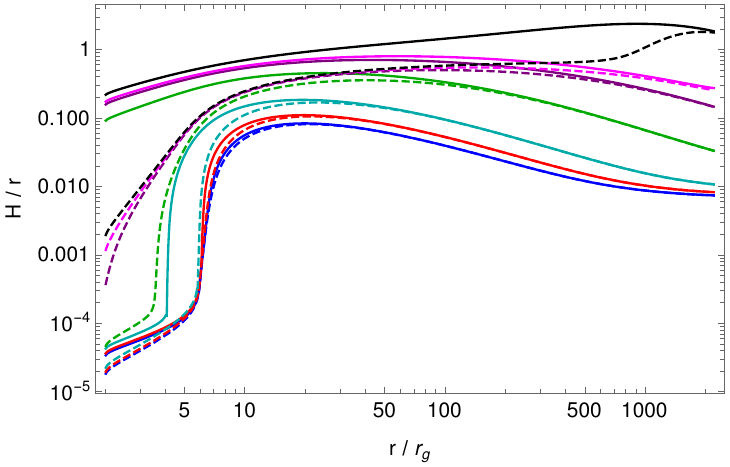}}
\subfigure[]{\includegraphics[scale = 0.67]{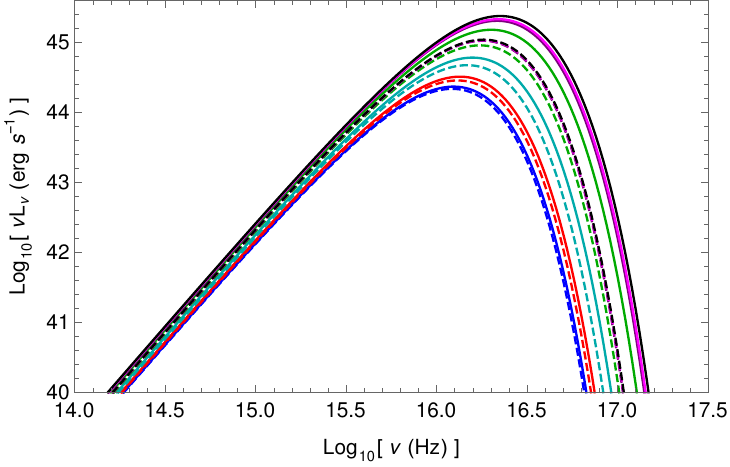}}
\caption{Radial profiles of the surface density (panel a), the scale-height-to-radius ratio (panel b), and the disk spectrum (panel c) for the $j=0$ case. The different colors represent different mass accretion rates. The solid and dashed lines denote Models III and IV, respectively.}
\label{fig:j0comp} 
\end{figure} 

\begin{figure}
\centering
\subfigure[]{\includegraphics[scale = 0.67]{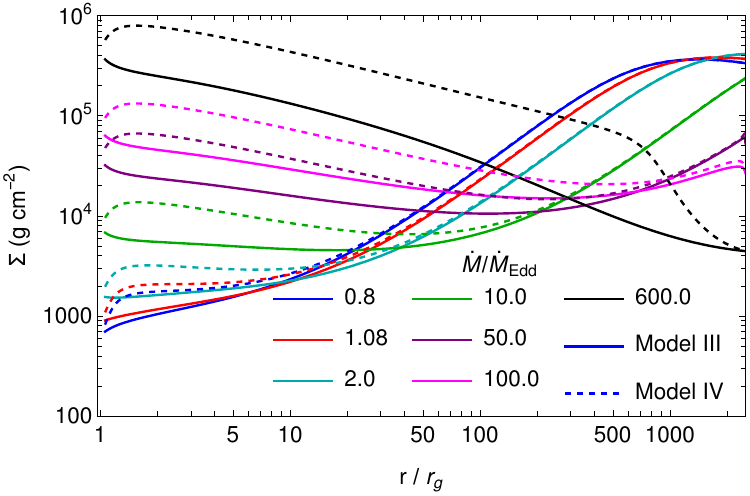}}
\subfigure[]{\includegraphics[scale = 0.67]{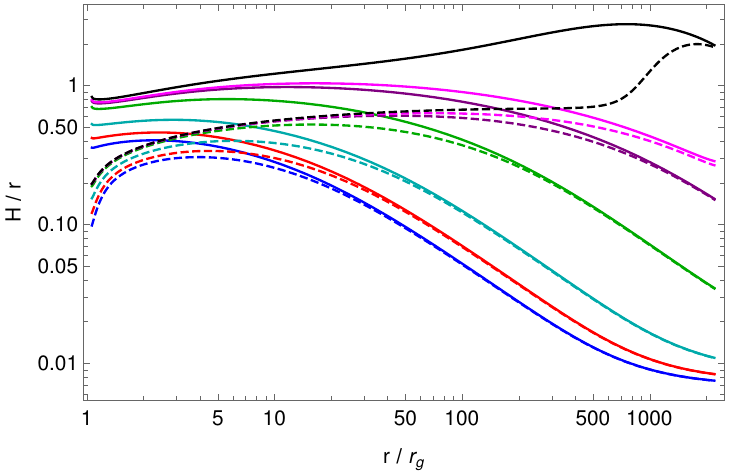}}
\subfigure[]{\includegraphics[scale = 0.67]{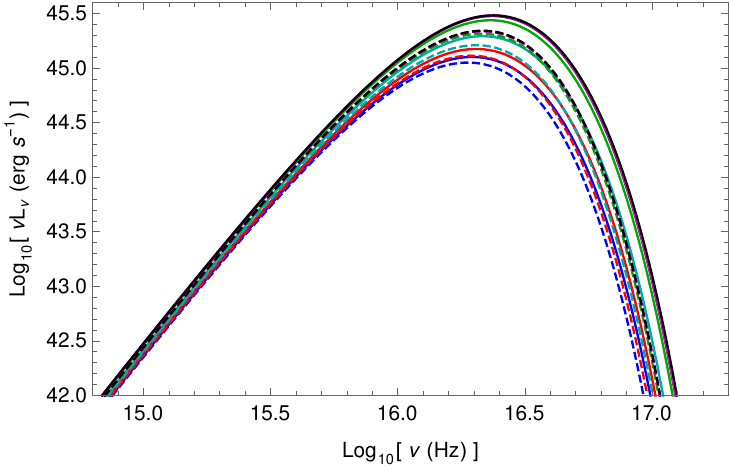}}
\caption{The same format as Figure~\ref{fig:j0comp} but for the $j=0.998$ case.}
\label{fig:j0998comp} 
\end{figure}

Figure~\ref{fig:j0comp} illustrates the radial dependence of the surface density, scale-height-to-radius ratio, and corresponding disk spectrum for the $j = 0$ case, while \ref{fig:j0998comp} has the same format but for the $j=0.998$ case. The scale height derivative affects the disk structure weakly at a low mass accretion rate for non-spinning black holes; however, the impact is significant for high-spinning black holes. This suggests that the increase in black hole spin increases the impact of scale height derivative even at the low mass accretion rate. The scale height derivative term affects the disk structure and spectrum crucially at $\dot{M}/\dot{M}_{\rm Edd} \gtrsim 2$ even for a non-spinning black hole. In the case of a highly spinning black hole, it has a significant influence on the disk structure near the ISCO radius, even for the sub-Eddington accretion rate.

\begin{figure*}
\centering
\subfigure[]{\includegraphics[scale = 0.73]{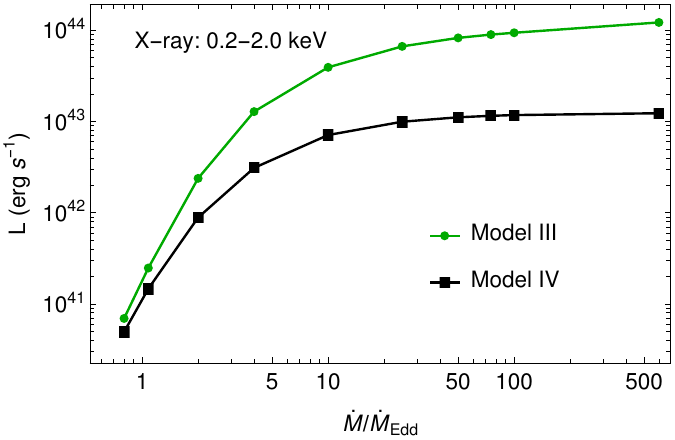}}
\subfigure[]{\includegraphics[scale = 0.72]{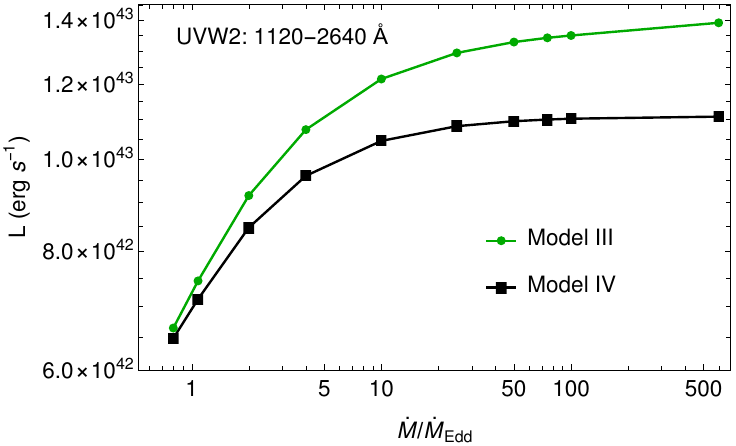}}
\caption{$\dot{M}$-dependence of X-ray (0.2-2 keV) and Swift UVW2 (1120-2640 \AA) luminosities for the $j=0$ case. The solid green and black lines represent each luminosity of Model III and Model IV, respectively. Each line is complemented between 10 data points of mass accretion rates.}
\label{fig:j0lumcomp} 
\end{figure*}

\begin{figure*}
\centering
\subfigure[]{\includegraphics[scale = 0.75]{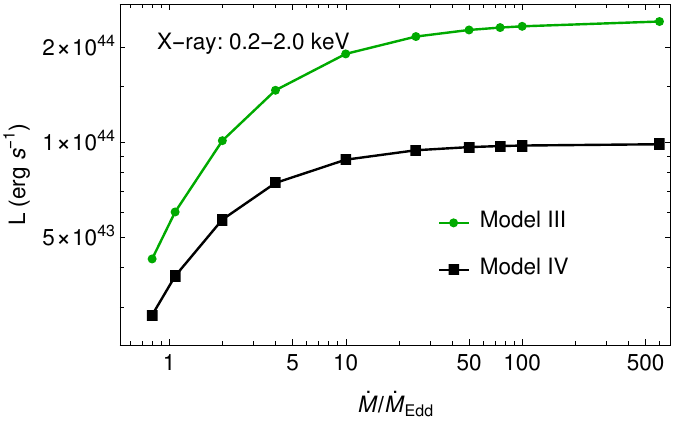}}
\subfigure[]{\includegraphics[scale = 0.75]{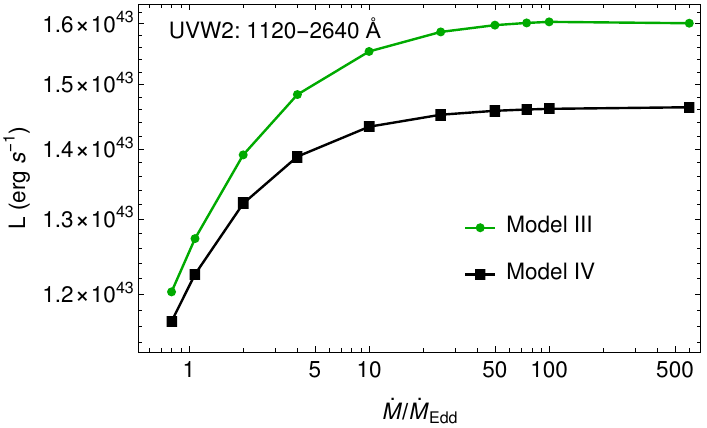}}
\caption{The same format as Figure~\ref{fig:j0lumcomp} but for the $j=0.998$ case.}
\label{fig:j0998lumcomp} 
\end{figure*}

The disk spectrum increases by including the scale height derivative, and the increment in the spectrum increases with the mass accretion rate. This results in the brightening of the disk spectrum in the soft X-ray waveband. The observed luminosities in the X-ray (0.2-2.0 keV) and Swift UVW2 (1120-2640 \AA) are given by $\displaystyle{L = 4 \pi D_{\rm L}^2 \int F_{\rm \nu, obs} \, \diff \nu_{\rm obs}}$, where equation (\ref{flux}) gives $F_{\rm \nu, obs}$. Figures \ref{fig:j0lumcomp} and \ref{fig:j0998lumcomp} depict $\dot{M}$-dependence of X-ray and UVW2 luminosities for both $j = 0$ and $0.998$, respectively. We note the increase in the X-ray luminosity due to the scale height derivative is much larger than that in the Swift UVW2 luminosity. The increment in the X-ray luminosity is twice or more than for $\dot{M}/\dot{M}_{\rm Edd}\gtrsim2$. Thus, the scale height derivative is more important for the soft-X-ray observations when $\dot{M}/\dot{M}_{\rm Edd} \gtrsim 2 $. While the luminosity of the $j=0.998$ case is somewhat larger than that of the $j=0$ case, the increment of the $j=0.998$ case is much smaller than that of the $j=0$ case. Notably, for the non-spinning BH case, the increment in the X-ray luminosity results in a one-order of magnitude difference for $\dot{M}/\dot{M}_{\rm Edd}\gtrsim100$.

%
\subsection{Importance of advective cooling for the entire disk}
\label{sec:advcooling}
%

For a radiation pressure-dominated slim disk, the ratio of advection to radiation cooling rates yields
\begin{equation}
\frac{Q_{\rm adv}}{Q_{\rm rad}} =\frac{1}{4} \frac{1}{\eta} \frac{\dot{M}}{\dot{M}_{\rm Edd}} \frac{r_{\rm g}}{r} \frac{H}{r} \xi(r),
\label{eq:qadvqrad}
\end{equation}
where equations (\ref{adveqn}) and (\ref{eq:anqrad}) with $\theta = 64$ are used for the derivation. Assuming that $\xi(r) \sim 1$ for simplicity, equation~(\ref{eq:qadvqrad}) means that the advective cooling rate is lower than the radiative cooling rate if $H/r \ll 1$ at $r \gg r_{\rm g}$. Moreover, this large/small relationship also holds even if $\dot{M} \gg \dot{M}_{\rm Edd}$ because $\dot{M} / \dot{M}_{\rm Edd}$ is comparable to $r/r_{\rm g}$. However, if $H/r \sim 1$ and $\dot{M} \gg \dot{M}_{\rm Edd}$, $Q_{\rm adv}$ is comparable with $Q_{\rm rad}$ even at $r \gg {r}_{\rm g}$. To confirm whether this condition is physically satisfied, we calculate H/r, by using equations~(\ref{eq:scalh})~and~(\ref{eq:anqrad}), to be
\begin{eqnarray}
\frac{H}{r} &=& \sqrt{2} \frac{c_{\rm s}}{c} \sqrt{\frac{r}{r_{\rm g}}} \nonumber \\
&=& \frac{3}{2}\frac{1}{\eta}\frac{\dot{M}}{\dot{M}_{\rm Edd}} \frac{r_{\rm g}}{r} \left[1 + \sqrt{1 +\frac{3}{4}\left(\frac{1}{\eta}\frac{\dot{M}}{\dot{M}_{\rm Edd}} \frac{r_{\rm g}}{r}\right)^2}\right]^{-1}
\label{eq:hrrad}
\end{eqnarray}
at the Keplerian rotation regime ($r \gg r_{\rm g}$). This equation indicates that $H/r \gtrsim 1$ at $r = 1000 r_{\rm g}$ for the $\dot{M} = 600 \dot{M}_{\rm Edd}$ case, yielding $Q_{\rm adv}/Q_{\rm rad} \sim 1$ from equation (\ref{eq:qadvqrad}). By substituting equation (\ref{eq:hrrad}) into equation (\ref{eq:qadvqrad}), $Q_{\rm adv}/Q_{\rm rad}$ can be more explicitly written as
\begin{multline}
\frac{Q_{\rm adv}}{Q_{\rm rad}} = \frac{3}{8} \xi(r) \left(\frac{1}{\eta}\frac{\dot{M}}{\dot{M}_{\rm Edd}} \frac{r_{\rm g}}{r}\right)^2 \times \\ \left[1 + \sqrt{1 + \frac{3}{4}\left(\frac{1}{\eta}\frac{\dot{M}}{\dot{M}_{\rm Edd}} \frac{r_{\rm g}}{r}\right)^2}\right]^{-1}.
\end{multline}
Above equation confirms that $Q_{\rm adv}/Q_{\rm rad} \gtrsim 1$ for $\eta = 0.1$, $\dot{M} \simeq (r/ r_{\rm g}) \dot{M}_{\rm Edd}$, and $\xi(r) \sim 1$, demonstrating that the advective cooling is important for the disk structure even at the far outer disk region if $\dot{M} \gg \dot{M}_{\rm Edd}$.

\begin{figure*}
\centering
\subfigure[$\dot{M} =1.08 \dot{M}_{\rm Edd}~{\rm and}~j = 0$]{\includegraphics[scale = 0.8]{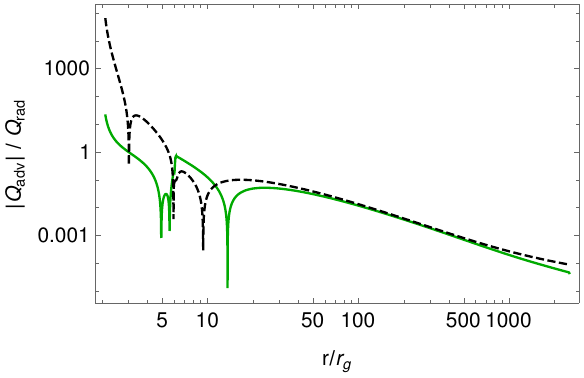}}
\subfigure[$\dot{M} =600 \dot{M}_{\rm Edd}~{\rm and}~j = 0$]{\includegraphics[scale = 0.8]{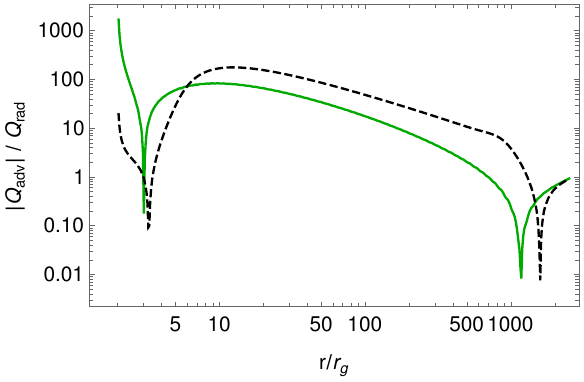}}
\subfigure[$\dot{M} =1.08 \dot{M}_{\rm Edd}~{\rm and}~j = 0.998$]{\includegraphics[scale = 0.8]{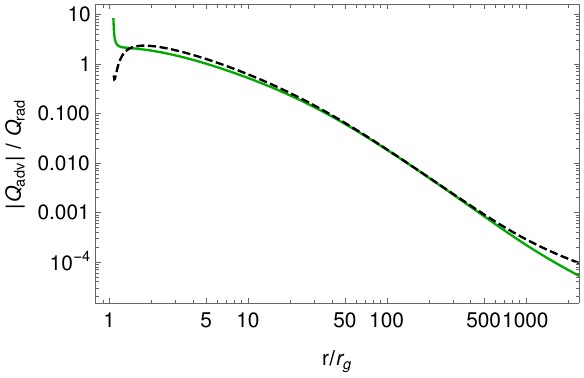}}
\subfigure[$\dot{M} =600 \dot{M}_{\rm Edd}~{\rm and}~j = 0.998$]{\includegraphics[scale = 0.8]{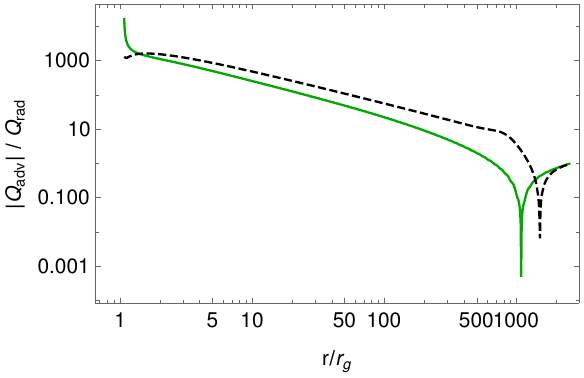}}
\caption{Radial dependence of the ratio of advective to radiative cooling rates, $|Q_{\rm adv}|/Q_{\rm rad}$, for $\dot{M} = 1.08 \dot{M}_{\rm Edd}~{\rm and}~600 \dot{M}_{\rm Edd}$. The upper and lower panels depict the $j=0$ and $j=0.998$ cases, respectively. The solid green and dashed black lines represent $|Q_{\rm adv}|/Q_{\rm rad}$ of Models III and IV, respectively.}
\label{fig:qratio}
\end{figure*}

This scaling presumption holds even if the $\xi(r) \sim 1$ assumption is relaxed. Figure~{\ref{fig:qratio}} the radial profile of the ratio of advective to radiative cooling rates. The solid green and dashed black lines represent $|Q_{\rm adv}|/Q_{\rm rad}$ of Models III and IV, respectively. The figure illustrates that $Q_{\rm adv}/Q_{\rm rad}\sim 1$ at $r\sim 1000\,r_{\rm g}$ when $\dot{M}=600 \dot{M}_{\rm Edd}$, while $Q_{\rm adv}/Q_{\rm rad}\ll 1$ at the same radius when $\dot{M}=1.08 \dot{M}_{\rm Edd}$. Note that the sharp droppings seen in panels (a), (b), and (d) of Figure \ref{fig:qratio} are caused by taking the absolute value of $Q_{\rm adv}$ at $Q_{\rm adv} = 0$ in the logarithmic scale. The zero value of $Q_{\rm adv}$ is due to a change from positive to negative values or vice versa (see also SA09). We confirm that advective cooling is crucial even for the disk's outer region if the mass accretion rate is much larger than the Eddington accretion rate.

%
\section{Discussion}
\label{discuss}
%

We have studied the impact of the angular momentum loss due to radiation and scale height derivative on a steady-state GR slim disk. We find that the term of the angular momentum loss due to the radiation $(r_{\rm g}^2/ c)x \ell Q_{\rm rad}$ is weaker than the term being proportional to the accretion rate $(1/2 \pi) \dot{M} c~ \diff \ell / \diff x$ so that becomes significantly less effective for higher mass accretion rate. In addition, because both the viscous heating and advective fluxes increase with the mass accretion rate, as can be seen from equations (\ref{viseqn}) and (\ref{adveqn}), the net radiation flux, which is given by $Q_{\rm rad} = Q_{\rm vis} - Q_{\rm adv}$, little increases with the mass accretion rate, especially in the super-Eddington regime. This also means the effect of the radiation-driven angular momentum loss also doesn't work effectively for a higher mass accretion rate.

Next, the impact of the scale height derivative on the disk structure is prominent in contrast to that of the radiation-driven angular momentum loss. The disk scale height increases due to the stronger radiation pressure at the more highly super-Eddington regime, resulting in a higher disk temperature. The stronger radiation pressure also decreases the surface density because the ratio of gas to total pressure, $\beta_{\rm gas}$, becomes lower. The disk scale-height derivative makes the surface density significantly lower. Note that the lower surface density implies that the radial disk velocity is higher if the mass accretion rate is constant. 

The black hole spin also impacts the disk structure mainly at the radius where the GR effects are significant and surely makes the disk structure deviate from the Newtonian one, although the mathematical dependence of the black hole spin on the surface density, disk temperature, and spectrum cannot be written explicitly as they are obtained numerically, unlike the NK solutions \citep{1973blho.conf..343N}. It is noted from Figures 1 and 5 that for $\dot{M}=1.08\,\dot{M}_{\rm Edd}$, the effective temperature rapidly decreases at the ISCO in the $j=0$ case. In contrast, it increases with the radius without any significant dropping in the case of $j=0.998$, extending the disk spectrum to a higher energy side. This difference is seen in the range of $\nu\gtrsim10^{16.5}$ of the disk spectrum in Figures 1d and 5d. It is noted from Figures 3 and 7 that for $\dot{M}=600\,\dot{M}_{\rm Edd}$, the surface density rapidly decreases around the ISCO in the $j=0$ case, whereas it increases with radius even beyond $r=2\,r_{\rm g}$ in the $j=0.998$ case. However, because the effective temperature around the inner edge radius doesn't change depending on the black hole spin, there is a slight difference in the resultant spectrum. Also, we find that these properties regarding the black hole spin effect on the disk structure show the same tendency between the model comprising the two key terms and the model without them.

Let us discuss how important the inclusion of the scale height derivative is for identifying parameters such as black hole mass and spins. Indeed, \citep{2022ApJ...933...31W} have statistically given a constraint on the black hole mass and spin by comparing the observed X-ray spectra of ASAS-SN 14li with both models of \citep{2020ApJ...897...80W,2021ApJ...918...46W}. Note that \citet{2020ApJ...897...80W} applied the GR disk model without including both the angular momentum loss due to radiation and the scale height derivative to the ASAS-SN 14li X-ray observations, while \citet{2021ApJ...918...46W} included only the radiation-driven angular momentum loss into the GR slim disk, which is consistent with our solutions for Models I and III. \cite{2021ApJ...918...46W} calculated the disk emission by using the photon ray-tracing method. The resultant X-ray spectra yield consistent results with those of \citet{2020ApJ...897...80W} because the radiation-driven angular momentum loss term has a significantly weak impact on the disk structure. Nonetheless, the model of \citet{2021ApJ...918...46W} provides a better $1\sigma$ contour than that of \citet{2020ApJ...897...80W}. \cite{2022ApJ...933...31W} have demonstrated that, unlike the case of \citet{2020ApJ...897...80W},  the parameters are well constrained within 1$\sigma$ contour when the radiation-driven angular momentum loss is included, even in spite of the weak impact on the disk structure and emission. As we described in Sections~\ref{sec:twoextreme} and \ref{sec:mdotdepend}, the scale-height derivative term is much more effective on the disk structure and emission than the radiation-derive angular momentum loss term. Therefore, we will make detailed calculations of the disk spectra by combining our GR model with the GR photon ray tracing code and then compare them with the TDE observations, making it possible to estimate the black hole mass and spin more precisely.

The flux sensitivity of the Swift X-ray telescope (XRT) is $f_{\rm lim} = 8 \times 10^{-14}~{\rm erg~s^{-1}~cm^{-2}}$. The luminosity distance of the ASAS-SN 14li is $D_{\rm L} = 90~{\rm Mpc}$ \citep{2016MNRAS.455.2918H}, such that luminosity detection limit for the ASAS-SN 14li is $L_{\rm lim} = 4 \pi D_{\rm L}^2 f_{\rm lim} = 7.75 \times 10^{40}~{\rm erg~s^{-1}} (D_{\rm L}/90\,{\rm Mpc})^2$. The X-ray luminosity shown in Figures \ref{fig:j0lumcomp} and \ref{fig:j0998lumcomp} is significantly higher than the detection limit of the telescope for both cases with and without scale-height derivative. In addition, the photometric accuracy of Swift XRT is $\sim10\%$, which implies that the X-ray luminosity of $\gtrsim 10^{41}~{\rm erg~s^{-1}}$ can be well measured.

In TDEs, the stellar debris is circularized to form an accretion disk with or without still infalling debris \citep{2020A&A...642A.111C}. If the accretion timescale is smaller than the fallback timescale of the infalling debris, we expect the mass accretion rate to follow the mass fallback rate $\dot{M}_{\rm fb}$. The luminosity will be then $L \propto \dot{M}_{\rm fb} \propto t^{n}$, where $n = -5/3$ for full TDEs and $n \simeq -9/4$ \citep{2019ApJ...883L..17C}. However, the TDE light curves in the soft X-ray waveband have shown different power-law evolution \citep{2017ApJ...838..149A} and indicates that the viscous accretion timescale of the TDE disk is usually longer than the fallback timescale. Because the fallback mass is likely to inject into the TDE disk from the outer edge at $\dot{M}_{\rm fb}$, there is an issue with how the mass accretion rate relates to the mass fallback rate and also depends on radius and time. In addition, according to the past literature on the non-GR and time-dependent ADAF models \citep{2002ApJ...576..908J,2006MNRAS.368..379M,2007ApJ...666..368L,2011ApJS..195....7X}, they have demonstrated the limit cycle behavior, which is caused by thermal instability, on the light curves. However, those models are assumed to be in a steady state at the initial time, and then the mass is supplied at the outer radius at a constant rate. This is clearly a different initial condition from that of TDEs. Therefore, there can be another problem with how the initial condition affects the limit cycle behavior. These questions motivate us to contract a time-dependent model of the optically thick ADAF in the TDE context in the near future.

%
\section{Conclusions}
\label{sac}
%

We have constructed a steady-state, optically thick, advection-dominated GR accretion disk with the alpha-viscosity prescription and both gas and radiation pressures by taking account of the angular momentum loss due to the radiation and scale-height derivative. Notably, the scale-height derivative term in the basic equations has hitherto been overlooked in extant literature (e.g. \cite{1994ASIC..417..341L,1996ApJ...471..762A,1998ApJ...498..313G,1998MNRAS.297..739B,2009ApJS..183..171S,2011A&A...527A..17S,2020ApJ...897...80W,2021ApJ...918...46W,2022ApJ...933...31W}).

We have studied the impact of these two physics on the disk structure and emission. Our primary conclusions are summarized as follows:
\begin{enumerate}
\item 
For comparison purposes, we have newly derived an analytical solution for a stationary radiation-pressure-dominant GR slim disk with zero viscous stress at the inner radius. 

\item 
The angular momentum loss due to radiation only slightly increases the surface density, disk temperature, and resultant spectrum near the disk's inner-edge radius. In addition, these effects become negligibly small as the mass accretion rate increases. This indicates that the radiation-driven angular momentum loss has little impact on the disk surface density, temperature, and spectrum for highly super-Eddington accretion flows.

\item
In contrast to the radiation-driven angular momentum loss case, the scale height derivative reduces the disk surface density while it increases the disk temperature and scale height. These effects significantly increase as the mass accretion rate increases, affecting the entire disk region at an extremely high accretion rate. 

%

%
\item 
The increment in the X-ray (0.2 - 2.0 keV) luminosity due to scale height derivative is significant for $\dot{M}/\dot{M}_{\rm Edd} \gtrsim 2 $. Moreover, the increment is higher for the non-spinning black hole than the spinning black hole case, resulting in a one-order of magnitude difference for $\dot{M}/\dot{M}_{\rm Edd}\gtrsim100$. These results indicate solving a set of basic equations with the scale height derivative for a super-Eddington accretion flow with $\dot{M}/\dot{M}_{\rm Edd} \gtrsim 2$ is crucial, especially for comparing with soft-X-ray observations.
\end{enumerate}

%
\begin{acknowledgments}
The authors have been supported by the Basic Science Research Program through the National Research Foundation of Korea (NRF) funded by the Ministry of Education (2016R1A5A1013277 to K.H. and 2020R1A2C1007219 to K.H. and M.T.). This work was financially supported by the Research Year of Chungbuk National University in 2021.
K.H. acknowledges the Institute for Theory and Computation, Harvard-Smithsonian Center for Astrophysics, for the warm hospitality during the sabbatical year. We thank the referee for the constructive suggestions that have improved the paper.
\end{acknowledgments}
%

\appendix

%
\section{Basic quantities of general relativistic disk equations}
\label{kmfv}
%

The Kerr metric in the Boyer-Lindquist coordinate is transformed to a cylindrical coordinate $\{t,~r,~\phi,~z\}$. The space-time metric in the geometrical units ($c=G =1$) with the signature ($- + + +$), is given by \citep{2020MNRAS.496.1784M}

\begin{widetext}
\begin{multline}
\diff S^2 = -\left[1-\frac{2 M (r^2+z^2)^{3/2}}{(r^2+z^2)^2+a^2 z^2}\right] \diff t^2 - \frac{4 M a r^2 \sqrt{r^2+z^2}}{(r^2+z^2)^2+a^2 z^2} \diff t \diff \phi + \frac{(r^2+z^2)^2+a^2 z^2}{(r^2+z^2)^2}\left[\frac{r^2}{r^2+z^2- 2 M \sqrt{r^2+z^2} +a^2}+ \right. \\ \left. \frac{z^2}{r^2+z^2}\right] \diff r^2 + \frac{(r^2+z^2)^2+a^2 z^2}{(r^2+z^2)^2} \left[\frac{z^2}{r^2+z^2- 2 M \sqrt{r^2+z^2} +a^2}+ \frac{r^2}{r^2+z^2}\right] \diff z^2 + \\ \frac{2 r z [(r^2+z^2)^2+a^2 z^2]}{(r^2+z^2)^2} \left[ \frac{1}{r^2+z^2- 2 M \sqrt{r^2+z^2} +a^2}- \frac{1}{r^2+z^2}\right] \diff r \diff z +  \frac{r^2}{r^2+z^2} \left[r^2+z^2+a^2 + \frac{2 M a r^2 \sqrt{r^2+z^2}}{(r^2+z^2)^2+a^2 z^2}\right] \diff \phi^2.
\end{multline}
\end{widetext}

In the limit of thin disc $z \ll r$, the metric tensors are given by 

\begingroup
\allowdisplaybreaks
\begin{eqnarray}
g_{\rm t t}&=& -1+ \frac{2 M }{r}- \frac{M (2 a^2+ r^2) z^2}{r^5}, \\ 
g_{\rm t r}&=& g_{\rm r t} = 0, \\
g_{\rm t \phi} &=& g_{\rm \phi t} =  -\frac{2 M a}{r} + \frac{M (2 a^3 + 3 a r^2)z^2}{r^5}, \\
g_{\rm t z} &=& g_{\rm z t} = 0, \\
g_{\rm r r} &=& \frac{r^2}{r^2 -2 M r + a^2} + \nonumber \\
&&  \frac{[2 a^4 + 3 a^2 (r^2-2M r)+ M (4 M r^2-3 r^3) ] z^2}{r^2(r^2- 2 M r +a^2)^2}, \nonumber \\
\label{grr} \\
g_{\rm r \phi} &=& g_{\rm \phi r} = 0, \\
g_{\rm r z} &=& g_{\rm z r} = 2 \left(-1+ \frac{r^2}{r^2 - 2 M r + a^2}\right) \frac{z}{r}, \label{gpp}\\
g_{\rm \phi \phi} &=& \frac{r^4+ a^2 r^2 + 2 M a^2 r}{r^2} - \nonumber \\ 
&&  \frac{a^2(r^4 + M r (2 a^2 + 5 r^2)) z^2}{r^6}, \\
g_{\rm \phi z} &=& g_{\rm z \phi} = 0, \\
g_{\rm z z} &=& 1 + \left[\frac{a^2- r^2}{r^2} + \frac{r^2}{r^2- 2 M r + a^2} \right] \frac{z^2}{r^2},
\end{eqnarray}
\endgroup

\noindent which is the same as the metric tensor given in \citet{2015PhyU...58..527Z}. At the equatorial plane ($z=0$), the space-time metric reduces to 

\begin{equation}
\diff S^2 = - \left(\frac{r- 2 M}{r}\right) \diff t^2 - \frac{4 M a }{r} \diff t \diff \phi + \frac{r^2}{\Delta} \diff r^2 + \frac{A}{r^2} \diff \phi^2 + \diff z^2,
\end{equation}

\noindent where $\Delta = r^2- 2 M r +a^2$ and $A= r^4+ a^2 r^2 + 2 M a^2 r$. In Boyer-Lindquist coordinates using orthonormal tetrad in the local non-rotating frame (LNRF) \citep{1972ApJ...178..347B}, the contravariant components of four velocities are given by \citep{1996ApJ...471..762A,2011ApJS..195....7X} 

\begin{eqnarray}
u^{\rm t} &=& \frac{\gamma_{\rm L} A^{1/2}}{r \Delta^{1/2}}, \\
u^{\rm r} &=& \frac{V}{\sqrt{1-V^2}} \frac{\Delta^{1/2}}{r}, \\
u^{\rm \phi} &=& \frac{r^2 \mathcal{L}}{A} + 2 M a \frac{\gamma_{\rm L} }{ A^{1/2} \Delta^{1/2}}, \\
u^{\rm z} &=& 0,
\end{eqnarray}

\noindent where $V$ is the radial velocity in the co-rotating frame, $\mathcal{L}$ is the angular momentum per unit mass, and $\gamma_{\rm L}$ is the Lorentz factor in LNRF near the equatorial plane given by

\begin{equation}
\gamma_{\rm L}^2 = \frac{1}{1-V^2} + \frac{r^2 \mathcal{L}^2}{A}.
\end{equation}

The covariant component of four velocities is given by

\begin{eqnarray}
u_{\rm t} &=& - \frac{\gamma_{\rm L} r \Delta^{1/2}}{A^{1/2}} - \frac{2 M a r}{A} \mathcal{L}, \label{utl}\\
u_{\rm r} &=&  \frac{r}{\Delta^{1/2}} \frac{V}{\sqrt{1-V^2}}, \label{url}\\
u_{\rm \phi} &=& \mathcal{L}, \label{upl}\\
u_{\rm z} &=& 0. \label{uzl}
\end{eqnarray} 

The viscous stress tensor is given by $S_{\beta}^{\alpha}= -2 \eta  \sigma_{\beta}^{\alpha}$, where $\sigma_{\beta}^{\alpha}= g^{\alpha i} \sigma_{i \beta}$, and $\sigma_{i \beta}$ is given by

\begin{multline}
\sigma_{i \beta} = \frac{1}{2} \left[\frac{\partial u_i}{\partial \mathbf{X}^{\beta}} + \frac{\partial u_\beta}{\partial \mathbf{X}^{i}} - 2 \Gamma^{\lambda}_{i \beta} u_{\lambda} + u^{\nu} \left(u_i u_{\beta}\right)_{; \nu} \right]- \\ \frac{1}{3} u^{\nu}_{; \nu} \left[g_{i \beta} + u_i u_{\beta}\right],
\label{sigeq}
\end{multline}

\noindent where $\Gamma$ is the Christoffel symbol and $\mathbf{X} \equiv \{t,~r,~\phi,~z\}$ represent four coordinates. For a subsonic flow such that angular velocity is smaller than the sound speed $\sim \sqrt{p/\rho}$, the $\sigma_{\phi}^r$ is given by

\begin{equation}
\sigma_{\phi}^r = \frac{1}{2} \frac{\Delta^{1/2} A^{3/2} \gamma_{\rm L}^3}{r^5} \frac{\diff \Omega}{\diff r},
\end{equation}  

\noindent where $\Omega = u^{\rm \phi}/u^{\rm t}$. The dynamic viscosity $\eta= \nu \rho$, where $\nu$ is the kinematic viscosity. The viscous stress is then given by $S_{\phi}^r = -2 \nu \rho \sigma_{\phi}^r$, and after vertical integration, it is given by 

\begin{equation}
\bar{S}_{\phi}^r = - \nu \Sigma \frac{\Delta^{1/2} A^{3/2} \gamma_{\rm L}^3}{r^5 } \frac{\diff \Omega}{\diff r},
\label{sphir}
\end{equation}

\noindent which is the same as the viscous stress obtained by \citet{1994ASIC..417..341L}. 

The mass conservation equation is given by $\left(\rho u^i\right)_{;i} = 0$, which after vertical integration, results in \citep{2011ApJS..195....7X}

\begin{equation}
\frac{\partial}{\partial t}\left(\Sigma u^t\right) + \frac{1}{r}\frac{\partial}{\partial r}\left(r \Sigma u^r\right) = 0,
\end{equation}

\noindent where $\Sigma = \int \rho \, \diff z = 2 H \rho$ and $H$ is the scale-height of the disk. For a steady disk, mass accretion rate $\dot{M} = -2 \pi r \Sigma u^r$ is a constant and is given by \citep{1996ApJ...471..762A}

\begin{equation}
\dot{M} = -2 \pi \Sigma \Delta^{1/2} \frac{V}{\sqrt{1-V^2}}. 
\label{mconswd}
\end{equation}

The radial conservation equation is given by $T^{ir}_{;i} = 0$, which results in \citep{2011ApJS..195....7X}

\begin{equation}
\frac{\partial V}{\partial t} = \frac{\Delta \sqrt{1-V^2}}{\gamma_{\rm L} A^{1/2}} \left[-\frac{V}{1-V^2} \frac{\partial V}{\partial r} + \frac{\mathcal{A}}{r} - \frac{1-V^2}{\rho} \frac{\partial p}{\partial r}\right],
\label{radmomeqn}
\end{equation}

\noindent where

\begin{equation}
\mathcal{A} = -\frac{M A}{r^3 \Delta \Omega_{\rm K}^{+}\Omega_{\rm K}^{-}} \frac{(\Omega - \Omega_{\rm K}^{+})(\Omega - \Omega_{\rm K}^{-})}{1 - \tilde{\Omega}^2\tilde{R}^2},
\label{mathcalA}
\end{equation}

\noindent with $\Omega_{\rm k}^{\pm} = \pm M / (r^{3/2} \pm a M^{1/2})$, $\tilde{\Omega} = \Omega - 2 M a r /A$, and $\tilde{R} = A / r^2 \Delta^{1/2}$. 
For the steady case and after the vertical integration, we have 

\begin{equation}
\frac{V}{1-V^2} \frac{\diff V}{\diff r} =  \frac{\mathcal{A}}{r} - \frac{1-V^2}{\Sigma} \frac{\diff P}{\diff r},
\label{consrad}
\end{equation}

\noindent where $P = \int p \, \diff z = 2 H p$.

The angular momentum conservation equation is given by $\displaystyle{\left(T^i_k \xi^k \right)_{; i}=0}$ \citep{2011ApJS..195....7X}, where $\xi^k \equiv \delta_{\,\phi}^k$ is the azimuthal Killing vector and $\delta_{\,\phi}^k$ is the Kronecker delta. It results in 

\begin{equation}
\rho u^t \frac{\partial u_{\phi}}{\partial t} + \rho u^r \frac{\partial u_{\phi}}{\partial r} + \frac{1}{r}\frac{\partial (r S^r_{\phi})}{\partial r} + \frac{\partial}{\partial z}(u_{\phi} q^z) = 0,
\end{equation}

\noindent and performing vertical integration, we get

\begin{equation}
\Sigma u^t \frac{\partial u_{\phi}}{\partial t} + \Sigma u^r \frac{\partial u_{\phi}}{\partial r} + \frac{1}{r}\frac{\partial (r \bar{S}^r_{\phi})}{\partial r} + u_{\phi} Q_{\rm rad} = 0, 
\end{equation}

\noindent where $Q_{\rm rad} = 2 q^z$ is the radiative flux. For a steady case, we get \citep{1996ApJ...471..762A}

\begin{equation}
\frac{\diff (r \bar{S}^r_{\phi})}{\diff r} = \frac{\dot{M}}{2 \pi} \frac{\diff \mathcal{L}}{\diff r} - r Q_{\rm rad} \mathcal{L}.
\label{eq:am}
\end{equation}

\noindent The viscous stress in the co-moving rotating frame obtained using the orthonormal tetrad basis is given by $t_{r\phi} = -r^2 \bar{S}_{\phi}^r / (\gamma_{\rm L} A^{1/2} \Delta^{1/2})$ \citep{1994ASIC..417..341L,1995ApJ...450..508R}. Assuming an alpha viscosity, $t_{r\phi} = -\alpha P$, we have 

\begin{eqnarray}
\bar{S}^r_{\phi} &=& \alpha P \frac{\gamma_{\rm L} A^{1/2} \Delta^{1/2}}{r^2}, \\
\nu &=& - \alpha \frac{P}{\Sigma} \frac{r^3}{\gamma_{\rm L}^2 A} \left(\frac{\diff \Omega}{\diff r}\right)^{-1} \nonumber \\
 &=& - 2\alpha \frac{H p}{\Sigma} \frac{r^3}{\gamma_{\rm L}^2 A} \left(\frac{\diff \Omega}{\diff r}\right)^{-1}.
\end{eqnarray}

The vertical hydrostatic equilibrium results in the scale height given by \citep{1997ApJ...479..179A}

\begin{equation}
\frac{P}{\Sigma H^2} = \frac{\mathcal{L}^2 - a^2 (\epsilon^2 -1)}{2 r^4} \equiv \zeta,
\label{heightwd}
\end{equation}

\noindent where $\epsilon = u_{\rm t} $ is the conserved energy for test particle motion.
 
The energy conservation equation is given by $Q_{\rm vis} = Q_{\rm rad} + Q_{\rm adv}$. Each term are written by \citep{2011ApJS..195....7X}

\begin{eqnarray}
Q_{\rm vis} &=& \nu \Sigma  \frac{\gamma_{\rm L}^4 A^2 }{r^6} \left(\frac{\diff \Omega}{\diff r}\right)^{2} = - \alpha P \frac{A \gamma_{\rm L}^2}{r^3} \frac{\diff \Omega}{\diff r}, 
\\
Q_{\rm rad} &=& \frac{\theta}{3}\frac{\sigma T^4}{\kappa \Sigma},
\label{qradwd}
\\
Q_{\rm adv} &=& \frac{1}{2\pi}\frac{\dot{M}}{r^2} \frac{P}{\Sigma}  \xi(r),
\end{eqnarray}

\noindent where $\theta = 64$ is adopted in our models and in those of \citet{2020ApJ...897...80W,2021ApJ...918...46W}, whereas $\theta = 32$ in SA09.
In addition, $\xi(r)$ is defined as 

\begin{eqnarray}
\xi(r) &\equiv& - \frac{T}{c_{\rm s}^2}\frac{\partial s }{ \partial \ln(r)} \nonumber \\
&=& -\frac{4 - 3 \beta_{\rm gas}}{\Gamma_3 - 1} \frac{\diff \ln T}{\diff \ln r} + (4 - 3 \beta_{\rm gas}) 
\biggr(
\frac{\diff \ln \Sigma}{\diff \ln r} - \frac{\diff \ln H}{\diff \ln r} 
\biggr),
\nonumber \\
\end{eqnarray}

\noindent where $s$ is the specific entropy, $\beta_{\rm gas} = p_{\rm gas} / p$ is the ratio of gas to total pressure, and $\Gamma_3$ is the third adiabatic exponent:

\begin{equation}
\Gamma_3 = 1+ \frac{(4 -3 \beta_{\rm gas})(\gamma_{\rm gas} -1)}{\beta_{\rm gas} + 12 (1-\beta_{\rm gas})(\gamma_{\rm gas}-1)}
\nonumber
\end{equation}

\noindent with the gaseous specific heat ratio, $\gamma_{\rm gas}$.

Employing the following dimensionless quantities $x = r/r_{\rm g},~~j = a/M,~~{\rm and}~~\displaystyle{\ell = \mathcal{L}/(r_{\rm g} c)}$, we obtain $A = r_{\rm g}^4 A_k$, $\Delta= r_{\rm g}^2 \Delta_k$, $\displaystyle{\Omega = (c/r_{\rm g})\omega}$, and 
$\gamma_{\rm L} =\sqrt{ 1/(1-V^2) + x^2\ell^2/A_k}$, where $A_k = x^4 + x^2 j^2 +2 x j^2$, $\Delta_k= x^2 -2 x +j^2$, and 
$\displaystyle{\omega = 2 j x/A_k + x^3 \Delta_k^{1/2} \ell/A_k^{3/2}}$, respectively. The conservation equations shown in Section~\ref{rmodel} are written in unit of these dimensionless variable.

%
\section{
Each component of $a_i$, $e_i$, and $r_i$
}
\label{compeqn}
%

We present the each component of $a_i$, $e_i$, and $r_i$ that appears in the equations (\ref{eq:vtd}), (\ref{eq:Ttd}) and (\ref{eq:ltd}) given in section \ref{sec:diskfull} to obtain the full GR slim disk solution. 

\begingroup
\allowdisplaybreaks
\begin{eqnarray}
r_1 &=& \frac{V^2}{1-V^2} - \frac{c_{\rm s}^2}{c^2} + 2 \frac{c_{\rm s}^2}{c^2} \left(\frac{1-\beta_{\rm gas}}{1+\beta_{\rm gas}} \right) \chi_2, \\
r_2 &=& 2 \frac{c_{\rm s}^2}{c^2}, \\
r_3 &=& 2 \frac{c_{\rm s}^2}{c^2} \left(\frac{1-\beta_{\rm gas}}{1+\beta_{\rm gas}} \right) \chi_3, \\
r_4 &=& \mathcal{A}_1 + \frac{c_{\rm s}^2}{c^2} \left[\frac{1}{2 \Delta_k} \frac{\diff \Delta_k}{\diff x} - 2 \left(\frac{1-\beta_{\rm gas}}{1+\beta_{\rm gas}} \right) \chi_4\right], 
\end{eqnarray}
\endgroup

\begingroup
\allowdisplaybreaks
\begin{eqnarray}
a_1 &=& 2 \chi_2 \left(\frac{1-\beta_{\rm gas}}{1+\beta_{\rm gas}} \right)- \frac{1}{\gamma_{\rm L}^2}\left\{1 + \frac{x^2 \ell^2}{A_k}\right\}, \\
a_2 &=& 2,\\
a_3 &=& \frac{x^2 \ell}{A_k \gamma_{\rm L}^2} - 2 \left(\frac{1-\beta_{\rm gas}}{1+\beta_{\rm gas}} \right) \chi_3 - \nonumber \\
&& \left\{
\begin{array}{ll}
\frac{1}{2\pi} \frac{\dot{M}c}{x \bar{S}_{\phi}^r} &~~~~ {\rm Model:~I~\&~II} \\
\frac{1}{\ell-\ell_{\rm in}}&~~~~ {\rm Model:~III~\&~IV} 
\end{array}
\right., \\
a_4 &=& -2 \left(\frac{1-\beta_{\rm gas}}{1+\beta_{\rm gas}}\right) \chi_4 + \frac{1}{2 \gamma_{\rm L}^2 (1-V^2)} \frac{\diff}{\diff x} \left[ \log\left(\frac{x^2}{A_k}\right)\right]\nonumber \\
&& - \left\{
\begin{array}{ll}
\frac{r_{\rm g}^2}{c} \frac{x \ell Q_{\rm rad}}{x \bar{S}^{r}_{\phi}} &~~~~ {\rm Model:~I~\&~II} \\
0 &~~~~ {\rm Model:~III~\&~IV} 
\end{array}
\right., 
\end{eqnarray}
\endgroup

\begingroup
\allowdisplaybreaks
\begin{eqnarray}
e_1 &=& \alpha \frac{ x^2 \ell V}{\gamma_{\rm L} A_k^{1/2} \sqrt{1-V^2}} -  \nonumber \\
&& \left\{
\begin{array}{ll}
2 \left(\frac{4-3 \beta_{\rm gas}}{1+\beta_{\rm gas}}\right)\chi_2 x &~~ {\rm Model:~I~\&~III} \\
(4-3 \beta_{\rm gas}) x &~~ {\rm Model:~II~\&~IV} 
\end{array}
\right., \\
e_2 &=& x \left[ \frac{1+\beta_{\rm gas}}{\Gamma_3 -1} + \right. \nonumber \\ 
&& \left. \left\{
\begin{array}{ll}
 4-3 \beta_{\rm gas} &~~ {\rm Model:~I~\&~III} \\
0 &~~ {\rm Model:~II~\&~IV} 
\end{array}
\right. \right],~~~~~~ \\
e_3 &=& - \alpha \frac{ x^2}{\gamma_{\rm L} A_k^{1/2} V \sqrt{1-V^2}} + \nonumber \\
&& \left\{
\begin{array}{ll}
2 \left(\frac{4-3 \beta_{\rm gas}}{1+\beta_{\rm gas}}\right)\chi_3 x &~~~~ {\rm Model:~I~\&~III} \\
0 &~~~~ {\rm Model:~II~\&~IV} 
\end{array}
\right., \\
e_4 &=&  \alpha \frac{\sqrt{1-V^2}}{V} \frac{\gamma_{\rm L}^2 A_k}{x \Delta_k^{1/2}} \chi_5 - 2 \pi \frac{r_{\rm g}^2 x^2}{\dot{M} c_{\rm s}^2} Q_{\rm rad} + \nonumber \\
&& \left\{
\begin{array}{ll}
2 \left(\frac{4-3 \beta_{\rm gas}}{1+\beta_{\rm gas}} \right) \chi_4 x &~~ {\rm Model:~I~\&~III} \\
(4-3 \beta_{\rm gas}) \frac{x}{2 \Delta_k} \frac{\diff \Delta_k}{\diff x} &~~ {\rm Model:~II~\&~IV} 
\end{array}
\right., 
\end{eqnarray}
\endgroup

\begingroup
\allowdisplaybreaks
\begin{eqnarray}
\chi_1 &=& \gamma_{\rm L} \frac{\diff}{\diff x}\left[\frac{x \Delta_k^{1/2}}{A_k^{1/2}}\right] + 2 j \ell \frac{\diff}{\diff x} \left[\frac{x}{A_k}\right] + \nonumber \\ 
&& \frac{1}{2}\frac{\ell^2}{\gamma_{\rm L}} \frac{x \Delta_k^{1/2}}{A_k^{1/2}}  \frac{\diff}{\diff x} \left(\frac{x^2}{A_k}\right),\\
\chi_2 &=& 1 +  \frac{1}{2} \frac{j^2 \epsilon}{\zeta(x) x^4} \frac{x \Delta_k^{1/2}}{A_k^{1/2}} \frac{V^2}{\gamma_{\rm L} (1-V^2)},\\ 
\chi_3 &=& \frac{1}{2}\frac{\ell - j^2 \epsilon \omega}{\zeta(x) x^4}, \\
\chi_4 &=& \frac{1}{2 \Delta_k} \frac{\diff \Delta_k}{\diff x} + \frac{1}{2}\frac{j^2 \epsilon \chi_1}{\zeta(x) x^4} + \frac{2}{x},\\
\chi_5 &=& 2 j \frac{\diff}{\diff x}\left(\frac{x}{A_k}\right) + \frac{\ell}{\gamma_{\rm L}} \frac{\diff}{\diff x}\left(\frac{x^3 \Delta_k^{1/2}}{A_k^{3/2}}\right)- \nonumber \\
&&\frac{1}{2} \frac{\ell^3}{\gamma_{\rm L}^3}  \frac{x^3 \Delta_k^{1/2}}{A_k^{3/2}} \frac{\diff}{\diff x}\left(\frac{x^2}{A_k}\right).
\end{eqnarray}
\endgroup

%
\section{Comparison with the earlier models}
\label{Work_comp}
%

\begin{figure*}
\centering
\subfigure[]{\includegraphics[scale = 0.65]{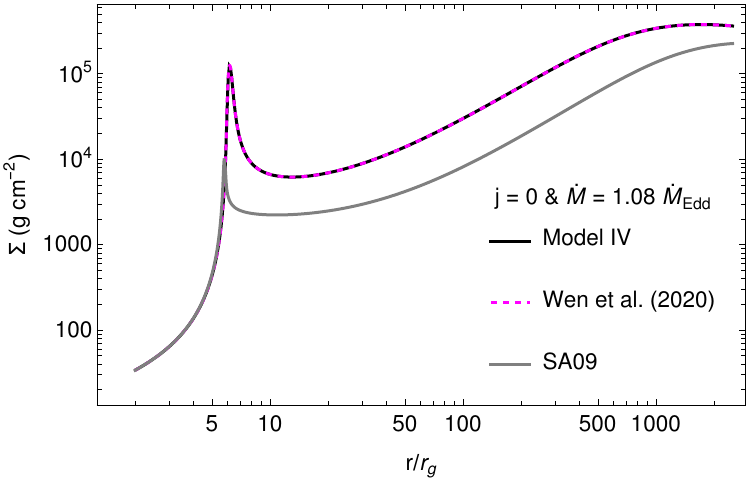}}
\subfigure[]{\includegraphics[scale = 0.65]{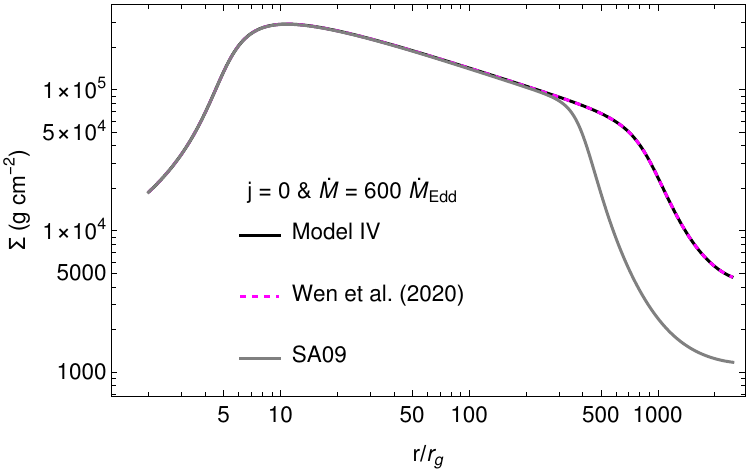}}
\subfigure[]{\includegraphics[scale = 0.65]{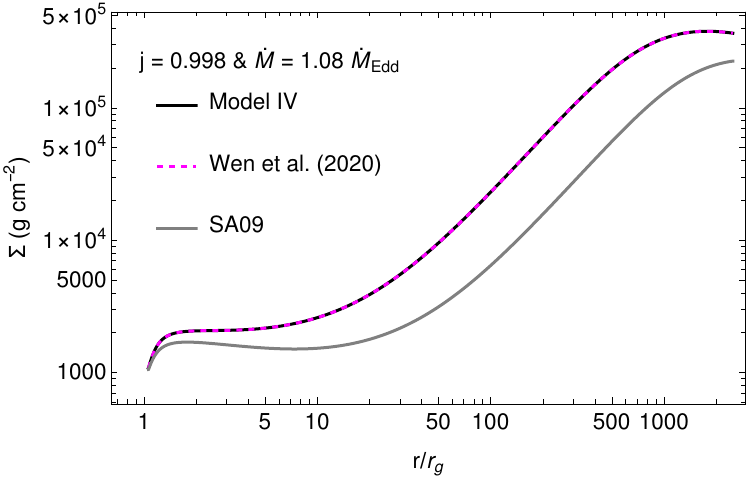}}
\subfigure[]{\includegraphics[scale = 0.65]{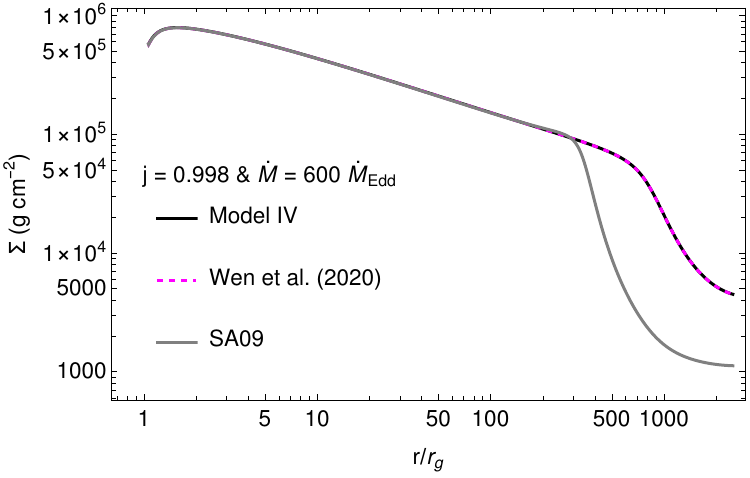}}
\caption{\label{sig_comp_IV}Comparison of the surface density's radial profiles between Model IV, the model of \citet{2020ApJ...897...80W}, and the SA09 model. Two different mass accretion rates $\dot{M} = 1.08 \dot{M}_{\rm Edd}$ and $\dot{M}=600 \dot{M}_{\rm Edd}$, and the two different black hole spins $j = 0$ and $0.998$ are adopted for each model.}
\end{figure*}

\begin{figure*}
\centering
\subfigure[]{\includegraphics[scale = 0.65]{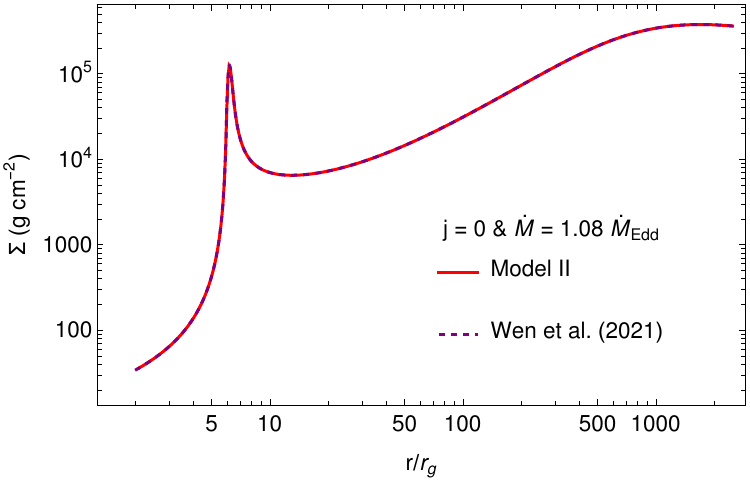}}
\subfigure[]{\includegraphics[scale = 0.65]{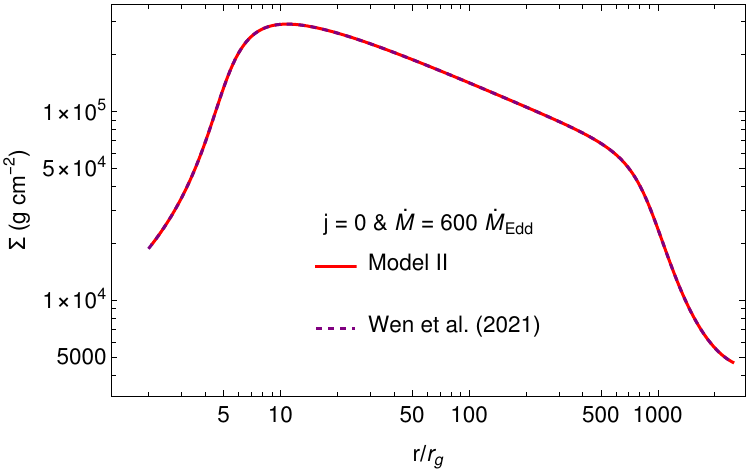}}
\subfigure[]{\includegraphics[scale = 0.65]{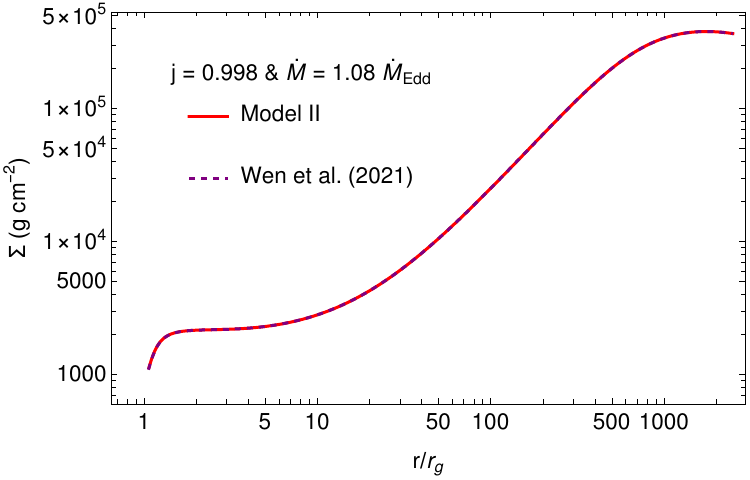}}
\subfigure[]{\includegraphics[scale = 0.65]{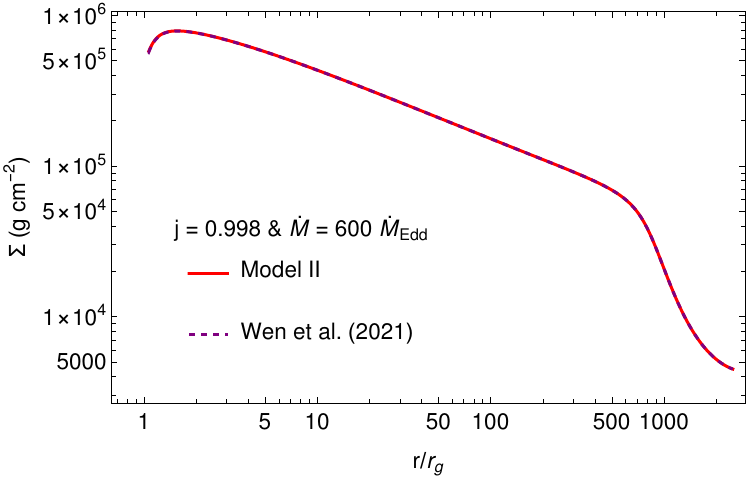}}
\caption{\label{sig_comp_II} Comparison of the surface density's radial profiles between Model II and the model of \citet{2021ApJ...918...46W}.  Two different mass accretion rates $\dot{M} = 1.08 \dot{M}_{\rm Edd}$ and $\dot{M}=600 \dot{M}_{\rm Edd}$, and the two different black hole spins $j = 0$ and $0.998$ are adopted for each model.}
\end{figure*}

In this section, we compare our models (Model II and Model IV) with the earlier three models (SA09, \citealt{2020ApJ...897...80W, 2021ApJ...918...46W}). The mass conservation equation (equation ~\ref{mconswd}) is commonly used among all the models. In the radial momentum conservation equation, SA09 and \citet{2020ApJ...897...80W, 2021ApJ...918...46W} neglect the $1-V^2$ term of the right-hand side of equation~\ref{consrad}, while our models include it. In the angular momentum conservation equation, Model II and the model of \citet{2021ApJ...918...46W} include the radiation-driven angular momentum term (i.e., the second term of the right-hand side of equation~\ref{eq:am}), while all the other models drop it. Regarding the energy equation, the value of the radiative cooling rate's coefficient, $\theta$, (see equation~\ref{qradwd}) is different between SA09 and the other models; $\theta=64$ is adopted for our models and \citet{2020ApJ...897...80W,2021ApJ...918...46W}, 
whereas $\theta=32$ for SA09.

Solving these basic equations numerically, we get the radial profiles of the surface density and other disk quantities for respective models. Figure~\ref{sig_comp_IV} compares the surface density's radial profile of Model IV with those of \citet{2020ApJ...897...80W}  and SA09, whereas Figure~\ref{sig_comp_II} compares the surface density's radial profile of Model II with that of \citet{2021ApJ...918...46W}. For all the models, we adopt the two different spin parameters $j=0$ and $j=0.998$ and the two different mass accretion rates $\dot{M} = 1.08 \dot{M}_{\rm Edd}$ and $600 \dot{M}_{\rm Edd}$. Note again that the three models seen in Figure~\ref{sig_comp_IV} include no radiation-driven angular momentum and scale-height derivative terms, whereas the two models appeared in Figure~\ref{sig_comp_II} include only the radiation-driven angular momentum loss.

From Figure~\ref{sig_comp_IV}, we find that the surface density of Model IV deviates from that of the SA09 model, while Model IV case is overlapped with the surface density profile of Wen et al.(2020). These behaviors are independent of the mass accretion rate and black hole spin, and are interpreted by using the simple analytical solutions as follows: Equation~(\ref{siganal}) demonstrates that the surface density is proportional to $(\theta / Q_{\rm rad})^2$, where $Q_{\rm rad}$ is also a function of $\theta$. Adopting $f(x,j)=1$ and $\zeta(x)=1/(2x^3)$, corresponding to the non-relativistic limit, in equation (\ref{eq:anqrad}), $Q_{\rm rad}$ is simply written as
\begin{equation}
Q_{\rm rad} \approx 
Q_{\rm vis} 
\left[
\frac{1}{2} + \sqrt{\frac{1}{4}+ 
\frac{1}{\eta^2}
\frac{768}{\theta^2} 
\left(
\frac{\dot{m}}{x}
\right)^2 
}
\right]^{-1},
\end{equation}
where $\dot{m}\equiv\dot{M}/\dot{M}_{\rm Edd}$. This equation shows that $Q_{\rm rad}$ is proportional to $\theta$ if $\dot{m}/x\gg1$, indicating the surface density is independent of $\theta$. This is consistent with the fact that, in the relevant disk region, the radial profiles of the surface density of Model IV is overlapped with those of the SA09 model. However, $Q_{\rm rad}$ is independent of $\theta$ if $\dot{m}/x\ll1$, indicating the surface density is proportional to $\theta^2$. This produces a significant (roughly factor 4) deviation in the radial profiles of the surface density between Model IV and the SA09 model.

Next, it is found from Figure~\ref{sig_comp_II} that there is no difference in the radial profiles of the surface density between Model II and the model of \citet{2021ApJ...918...46W}, even though there are a few differences in the basic equations between them. This means the $(1-V^2)$ term of the right-hand side of equation (\ref{consrad}) has effectively no impact on the disk structure.

\bibliography{reference}

\end{document}